\documentclass[11pt, a4paper]{article}
\pdfoutput=1
\usepackage[utf8]{inputenc}
\usepackage{jheppub}
\hypersetup{pdfencoding=unicode, bookmarksopen=true, bookmarksnumbered}
\usepackage{microtype}
\usepackage{amsthm,amsbsy,amsfonts,mathrsfs,enumerate,float,wrapfig,amsmath,mathtools,graphicx,framed,tcolorbox}
\usepackage{tikz}
\usepackage{subcaption}
\usepackage{graphbox}
\usepackage{todonotes}

\DeclareMathOperator{\PE}{PE}

\DeclareMathOperator{\Tr}{Tr}
\DeclareMathOperator{\tr}{tr}
\DeclareMathOperator{\rk}{rk}
\DeclareMathOperator{\sgn}{sgn}
\newcommand{\SO}[1]{\mathrm{SO}{(#1)}}
\newcommand{\so}[1]{\mathfrak{so}{(#1)}}
\newcommand{\SU}[1]{\mathrm{SU}{(#1)}}
\newcommand{\su}[1]{\mathfrak{su}{(#1)}}
\newcommand{\U}[1]{\mathrm{U}{(#1)}}

\newcommand{\thtftn}[2]{\theta\genfrac[]{0pt}{1}{#1}{#2}}

\usepackage{tikz}
\usetikzlibrary{positioning}
\usetikzlibrary{arrows}

\title{Quantum curves of 6d SCFTs on a -3 curve}

\author[a]{Minsung Kim,}
\author[b]{Sung-Soo Kim,}
\author[c,d]{Xin Wang}
\affiliation[a]{Quantum Universe Center, Korea Institute for Advanced Study, Seoul 02455, Korea}
\affiliation[b]{School of Physics, University of Electronic Science and Technology of China, Chengdu, Sichuan 611731, China}
\affiliation[c]{Interdisciplinary Center for Theoretical Study, University of Science and Technology of China, Hefei, Anhui 230026, China}
\affiliation[d]{Peng Huanwu Center for Fundamental Theory, Hefei, Anhui 230026, China}

\emailAdd{minsung1@kias.re.kr}
\emailAdd{sungsoo.kim@uestc.edu.cn}
\emailAdd{wxin@ustc.edu.cn}

\abstract{
    We study half-BPS defects and quantum Seiberg--Witten curves in 6d $ \mathcal{N}=(1,0) $ SCFTs on a $ -3 $ curve. We compute the partition functions in the presence of codimension-4 and codimension-2 defects on $ \mathbb{R}^4 \times T^2 $ using generalized blowup equations. We then propose quantum Seiberg--Witten curves as elliptic difference equations satisfied by the Nekrasov--Shatashvili limits of the defect partition functions. We apply this construction to the minimal $ \su 3 $ gauge theory, and to the $ \mathfrak{g}_2 $ and $ \so N $ gauge theories on a Higgsing tree whose endpoint is the minimal $ \su 3 $ theory. We also investigate defects and the quantum curve of the $ \su 3 $ gauge theory compactified on a circle with a $ \mathbb{Z}_2 $ twist.
}

\begin{document}
\preprint{\shortstack[r]{KIAS-Q26032\\USTC-ICTS/PCFT-26-76}}
\maketitle

\section{Introduction}

Seiberg--Witten (SW) geometry describes the exact low-energy dynamics of 4d $\mathcal{N}=2$ theories on the Coulomb branch \cite{Seiberg:1994rs,Seiberg:1994aj}. It consists of a family of curves and a meromorphic differential whose periods determine the effective gauge couplings and the central charges of BPS states. SW curves can also be constructed using integrable systems and geometric engineering \cite{Donagi:1995cf,Katz:1996fh}, and their study has led to the discovery of new 4d $\mathcal{N}=2$ superconformal theories \cite{Argyres:1995jj,Minahan:1996fg}. For 5d theories compactified on a circle, the curves contain contributions from Kaluza--Klein modes and are naturally written in multiplicative spectral coordinates. Such curves arise, for example, as spectral curves of relativistic integrable systems \cite{Nekrasov:1996cz,Brandhuber:1997ua}.

For 6d $\mathcal{N}=(1,0)$ superconformal field theories (SCFTs) compactified on $T^2$, SW geometry describes the low-energy dynamics of the resulting 4d $\mathcal{N}=2$ theories. The parameters of the geometry include the complex structure of $T^2$, flavor holonomies, and Coulomb parameters descending from the tensor and gauge multiplets. Examples include the mass-deformed E-string curve \cite{Ganor:1996xd,Eguchi:2002fc,Eguchi:2002nx} and the curves derived from the thermodynamic limit of string partition functions for ADE string chains and D-type theories \cite{Haghighat:2017vch,Haghighat:2018dwe}. More recently, SW curves have been proposed for little string theories \cite{Filoche:2024knd, Filoche:2024vne}. SW geometry has also been used in the classification of 6d SCFTs through their torus compactifications~\cite{Xie:2022lcm,Xie:2022aad,Xie:2023wqx,Xie:2025jhj}.

The SW geometry can be quantized by considering the theory on an $\Omega$-background with equivariant parameters $\epsilon_1$ and $\epsilon_2$. In the limit $\epsilon_1,\epsilon_2\to0$, the classical SW geometry is recovered, and its prepotential can be extracted from the leading asymptotics of the supersymmetric partition function \cite{Nekrasov:2002qd, Nekrasov:2003rj}. The Nekrasov--Shatashvili (NS) limit is instead obtained by taking $\epsilon_2\to0$ with $\epsilon_1=\epsilon$ fixed, where $\epsilon$ plays the role of a quantization parameter \cite{Nekrasov:2009rc}. In this limit, the SW curve is promoted to a quantum curve, expressed as a differential or difference equation. Its wavefunction can be obtained from an appropriately normalized partition function in the presence of a codimension-2 defect, as studied in gauge theory and refined topological strings \cite{Alday:2009fs,Kozcaz:2010af,Aganagic:2011mi, Gaiotto:2014ina, Bullimore:2014awa}.

For 6d theories, elliptic quantum curves have been constructed for theories associated with class $\mathcal{S}_k$ \cite{Chen:2020jla}, the E-string theory \cite{Chen:2021ivd}, and $\mathfrak{so}(N)$ or $\mathfrak{sp}(N)$ gauge theories with fundamental hypermultiplets \cite{Chen:2021rek,Chen:2023aet}. In these constructions, the wavefunction is obtained from the partition function of a codimension-2 defect on $\mathbb{C}_{\epsilon_1}\times T^2$, while codimension-4 defects wrapping $T^2$ determine Wilson surface observables appearing in the difference equation. The E-string and D-type equations are related to the van Diejen operator and elliptic Garnier systems, respectively \cite{Nazzal:2018brc, Nazzal:2021tiu}. Further quantum curves, including those for twisted compactifications, can be obtained by Higgsing \cite{Chen:2021rek,Chen:2023aet}. See also \cite{Gaiotto:2012xa, Bullimore:2014upa, Bullimore:2014awa, Razamat:2018zel, Hatsuda:2018lnv, Nazzal:2023wtw, Kim:2024mnp, Zeev:2026alb} for alternative approaches to the relation between 6d theories and integrable systems.

To construct quantum curves in this way, one needs to compute the bulk and defect partition functions. Given a quiver gauge theory description of the self-dual strings in a 6d theory with or without defects, the corresponding partition functions can be computed explicitly using localization techniques. However, the $\mathfrak{so}(N)$ theories on a $-3$ curve contain spinor matter, and for most of these theories no explicit worldsheet gauge theory description of the self-dual strings is known. The previous method therefore cannot be applied directly. It is thus interesting to ask how to obtain quantum curves for this class of theories and better understand their SW geometry.

In this paper, we study quantum SW curves for a family of 6d SCFTs supported on a single $-3$ curve in the F-theory base \cite{Heckman:2015bfa}. We consider the minimal theory with pure $\mathfrak{su}(3)$ gauge algebra, the $\mathfrak{g}_2$ theory with one fundamental hypermultiplet, and the $\mathfrak{so}(N)$ theories for $7\leq N\leq10$ with certain vector and spinor matter. These theories lie on a Higgsing tree whose endpoint is the non-Higgsable pure $\mathfrak{su}(3)$ theory. We also study the $\mathbb{Z}_2$ outer-automorphism twisted compactification of the pure $\mathfrak{su}(3)$ theory. To compute the bulk partition functions, we use either the ADHM-like constructions of \cite{Kim:2016foj,Kim:2018gjo} or elliptic blowup equations \cite{Gu:2018gmy,Gu:2020fem,Kim:2020hhh}. We also use their extensions to Wilson loops and defects \cite{Kim:2021gyj,Kim:2025qaf}. For codimension-2 defects and Wilson surfaces, we use proposed blowup equations that extend the approach of \cite{Kim:2025qaf} to theories with nontrivial gauge dynamics on the tensor branch.

Using these defect partition functions, we propose elliptic quantum SW curves for the theories considered above. Defining the shift operator $Y$ by $Y^n f(x)=f(x+n\epsilon)$, we write the quantum curves in the common form
\begin{align}\label{eq:so-curve-intro}
    \left( Y - A(x)Y^{-1} - B(x)Y^{-2} \right)\psi(x)
    = \chi(x)\psi(x) \, .
\end{align}
Here, $\psi(x)$ is the NS limit of the normalized codimension-2 defect partition function. The coefficient $A(x)$ is a linear combination of Wilson surface expectation values in the NS limit, while $B(x)$ is an explicit theta-function expression independent of the dynamical parameters. The function $\chi(x)$ is determined by the normalized NS expectation values of codimension-4 defects, which can be expressed as a combination of Wilson surface expectation values. The explicit quantum curves for the $\mathfrak{so}(N)$, $\mathfrak{g}_2$, and pure $\mathfrak{su}(3)$ theories are given in \eqref{eq:so-curve}, \eqref{eq:G2-curve}, and \eqref{eq:su3-curve}, respectively. We have checked these difference equations up to 3-string order.

We also study the Higgsing relations among these quantum curves. We consider the Higgsing along $\mathfrak{so}(10)\to\mathfrak{so}(9)\to\cdots\to \mathfrak{so}(7)\to\mathfrak{g}_2\to\mathfrak{su}(3)$. Applying these limits to the bulk and defect partition functions, we recover the quantum curves of the Higgsed theories. Higgsing with Kaluza--Klein momentum further relates the $\mathfrak{g}_2$ theory to the twisted $\mathfrak{su}(3)$ compactification \cite{Kim:2019dqn,Kim:2021cua}. For the twisted theory, we construct the defect partition functions and propose a quantum curve of the same form as \eqref{eq:so-curve-intro}. Its perturbative Wilson surface expectation values are given by characters of the twisted affine algebra $A_2^{(2)}$. The curve agrees with the Higgsing limit of the $\mathfrak{g}_2$ curve, and we have checked its difference equation up to 3-string order.

The paper is organized as follows. In section~\ref{sec:SWgeometryAndDefects}, we review the bulk partition functions and introduce codimension-2 and codimension-4 defects, their anomaly constraints, and the blowup equations. In section~\ref{sec:su3}, we study the pure $\mathfrak{su}(3)$ theory, compute its Wilson surface expectation values, and propose its quantum curve. We also compute the classical periods. In section~\ref{sec:higgsable}, we study the $\mathfrak{so}(N)$ and $\mathfrak{g}_2$ theories, their Higgsing relations, and the twisted $\mathfrak{su}(3)$ compactification. Our conventions for special functions and Lie algebras are summarized in appendices~\ref{app:special} and~\ref{app:Lie}.

\section{Seiberg--Witten geometry and defects in 6d SCFTs}\label{sec:SWgeometryAndDefects}

In this section, we review the SW curves and half-BPS defects in 6d SCFTs. We begin with classical and quantum SW geometry and its relation to partition functions. After introducing the partition functions on $ \Omega $-deformed $ \mathbb{R}^4 \times T^2 $, we discuss codimension-4 and codimension-2 defects, and their partition functions.

\subsection{Seiberg--Witten curve of 6d SCFTs}\label{sec:review}
The Seiberg--Witten (SW) geometry associated with a six-dimensional $\mathcal{N} = (1,0)$ superconformal theory (SCFT) describes the low-energy effective theory obtained by compactification on a two-torus $T^2$ \cite{Ohmori:2015pua}. With supersymmetry preserving boundary conditions, this compactification gives a four-dimensional $\mathcal{N}=2$ theory whose Coulomb branch effective dynamics is encoded by a family of algebraic curves, together with a meromorphic differential $\lambda_{\rm SW}$. The Coulomb parameters originate from both the six-dimensional tensor multiplets and the gauge multiplets on the tensor branch. Periods of $\lambda_{\rm SW}$ determine the special coordinates and their magnetic duals, and hence the effective prepotential. The resulting geometry incorporates not only the perturbative tensor-branch spectrum but also the nonperturbative contributions of self-dual strings. In suitable spectral presentations, the curve takes the form 
\begin{align}
    H(x, y ; u, m, \tau)=0, \quad 
\lambda_{\rm SW} = \log y dx\ ,
\end{align}
up to conventional normalizations and background-dependent choices of the differential. Here $\tau$ is the complex structure of the compactified torus $T^2$, $m$ denotes flavor parameters, and $u$ parametrizes the family of the curve. The spectral coordinate $x$ is elliptic. The coefficients of $H$ are expressed through theta functions or, more generally, meromorphic Jacobi sections \cite{Chen:2020jla,Chen:2021ivd,Chen:2021rek,Chen:2023aet}. 

The SW geometry also admits a description in terms of protected partition functions on $\Omega$-deformed $\mathbb{R}^4_{\epsilon_1, \epsilon_2}\times T^2$. The bulk partition function contains contributions from self-dual strings, organized into an expansion whose coefficients are their elliptic genera. In the limit where $\epsilon_1, \epsilon_2\to0$, the partition function determines the genus-zero free energy \cite{Nekrasov:2002qd, Nekrasov:2003rj},
\begin{align}
    \mathcal{F}_0 = \lim_{\epsilon_1, \epsilon_2\to0} \epsilon_1 \epsilon_2\log Z\ , 
\end{align}
which must agree with the prepotential obtained from the periods of the SW differential. Keeping one deformation parameter finite instead gives the Nekrasov--Shatashvili (NS) limit, which provides a quantization of the classical geometry \cite{Nekrasov:2009rc}. These two limits therefore connect the same protected partition function data to classical special geometry and to a quantum spectral problem, respectively.

Supersymmetric defects provide a particularly useful realization of the quantum spectral problem. A codimension-2 defect supported on $\mathbb{C}_{\epsilon_1}\times T^2$ introduces an additional parameter $x$, interpreted as a defect flavor holonomy and subsequently as the spectral coordinate. A codimension-4 defect, on the other hand, wraps $T^2$ and is localized at the origin of $\mathbb R^4$. By denoting the partition function with codimension-2 and codimension-4 by  $\Psi(x)$ and $\mathcal X(x)$, their normalized Nekrasov--Shatashvili limits define 
\begin{align}
\psi(x)= \left.\lim_{\epsilon_2\to0} \frac{\Psi(x-\epsilon_1/2)}{Z}\right|_{\epsilon_1=\epsilon}, \qquad \chi(x)= \left.\lim_{\epsilon_2\to0} \frac{\mathcal X(x)}{Z}\right|_{\epsilon_1=\epsilon}.    
\end{align}
With the appropriate classical and perturbative normalization, 
$\Psi(x)$ serves as a wavefunction satisfying an elliptic difference equation \cite{Aganagic:2011mi, Bullimore:2014awa}. Such constructions have been established for E-string, orthogonal, and D-type conformal-matter theories \cite{Haghighat:2018dwe, Chen:2021ivd,Chen:2021rek,Chen:2023aet}.

The normalized codimension-two partition function serves as a wavefunction for an elliptic difference operator. Introducing
\begin{align}
Yf(x)=f(x+\epsilon), \qquad X=e^{2\pi ix}, \qquad YX=e^{2\pi i\epsilon}XY,    
\end{align}
the quantum SW curve takes the schematic form
\begin{align}
\widehat H(x,Y; U)\,\psi(x)=0.   
\end{align}
The quantities $U$ are normalized NS-limit Wilson-surface expectation values. Codimension-four observables act as generating functions for these quantities, with elliptic coefficient functions carrying the dependence on $x$. Importantly, the operator need not be a second order eigenvalue equation: several shifts and nonlinear combinations of Wilson observables can occur. Its classical relation follows by taking $\epsilon\to 0$ and replacing shifts by the corresponding commuting spectral variable.

\subsection{Partition functions of 6d SCFTs}

In this subsection, we briefly review the partition functions of 6d SCFTs on the tensor branch and the blowup equations that determine them. The partition function of a 6d SCFT on the $ \Omega $-deformed $ \mathbb{R}^4\times T^2 $ is defined as
\begin{align}\label{eq:Z}
    Z = \Tr\left[ (-1)^F e^{2\pi i(\tau H_L - \bar\tau H_R)} e^{2\pi i \epsilon_1(J_1+J_R)} e^{2\pi i \epsilon_2(J_2+J_R)} e^{2\pi i (\langle \Pi, \phi \rangle + \langle F, m \rangle)} \right] \, ,
\end{align}
where $ \tau $ is the complex structure of the torus $ T^2 $, $ H_L $ and $ H_R $ are the left- and right-moving Hamiltonians, $ J_{1,2} $ and $ J_R $ are the Cartan generators of the $ \SO 4 \simeq \SU{2}_l \times \SU{2}_r $ Lorentz rotation of $ \mathbb{R}^4 $ and the $ \SU 2_R $ R-symmetry, respectively, $ \epsilon_{1,2} $ are the $ \Omega $-deformation parameters, and $ (\Pi, F) $ and $ (\phi, m) $ denote the gauge and flavor charges and the corresponding chemical potentials, respectively. The charges $ \Pi $ also include the tensor charges of the self-dual strings. This partition function counts the Ramond sector BPS spectrum and is independent of $ \bar\tau $.

Let us first describe the classical contribution to the partition function \eqref{eq:Z}. At a generic point of the tensor branch of a 6d SCFT, the bosonic part of the tree-level low energy effective action is given by \cite{Bonetti:2011mw, Bonetti:2013ela, Grimm:2015zea, Bhardwaj:2019fzv, Kim:2020hhh}
\begin{align}\label{eq:effS-tree}
    S = \int \left( -\frac{1}{2} \Omega^{\alpha\beta} G_{\alpha} \wedge *G_\beta - \Omega^{\alpha\beta} B_\alpha \wedge X_{4\beta} \right) \, ,
\end{align}
where $ \Omega^{\alpha\beta} $ is the negative-definite, symmetric bilinear form on the tensor charge lattice, $ B_\alpha $ are the self-dual 2-form tensor fields and $ G_\alpha $ are their field strengths. The second term in the effective action \eqref{eq:effS-tree} is called the Green--Schwarz term, and it cancels the 1-loop gauge anomaly via the Green--Schwarz--Sagnotti mechanism \cite{Sagnotti:1992qw}. The 4-form $ X_{4\alpha} $ is given by
\begin{align}\label{eq:X4}
    dG_\alpha = X_{4\alpha} \, , \quad
    X_{4\alpha} = -\frac{1}{4}a_\alpha p_1(T_6) + \frac{1}{4} \sum_a b_{a\alpha} \Tr F_a^2 + c_\alpha c_2(R) \, ,
\end{align}
where $ p_1(T_6) $ and $ c_2(R) $ are the first Pontryagin class of the tangent bundle of the 6d spacetime and the second Chern class of the $ \SU{2}_R $ R-symmetry bundle, respectively, and $ F_a $ is the field strength of the $ a $-th symmetry, where the index $ a $ runs over all gauge and flavor symmetries of the theory. The trace $ \Tr $ is normalized such that the instanton number $ \frac{1}{4}\Tr F_a^2 $ is an integer. The Green--Schwarz term contributes to the anomaly 8-form of the 6d theory as
\begin{align}
    I_{\mathrm{GS}} = -\frac{1}{2} \Omega^{\alpha\beta} X_{4\alpha} X_{4\beta} \, ,
\end{align}
and the coefficients $ a_\alpha $, $ b_{a \alpha} $ and $ c_\alpha $ are determined from the anomaly cancellation conditions \cite{Ohmori:2014kda}. Upon compactification on a circle, the 6d theory becomes a 5d Kaluza-Klein (KK) theory, and the tree-level action contributes to its effective prepotential on the $ \Omega $-background as \cite{Bhardwaj:2019fzv, Kim:2020hhh}
\begin{align}
    \epsilon_1 \epsilon_2 \mathcal{E}_0^{\mathrm{tree}}
    &= -\frac{\tau}{2} \Omega^{\alpha\beta} \varphi_{\alpha} \varphi_{\beta} - \Omega^{\alpha \beta} \varphi_{\alpha} \left( \frac{a_\beta}{4}(\epsilon_1^2+\epsilon_2^2) + \sum_a \frac{b_{a\beta}}{2} \langle \phi_a, \phi_a \rangle + c_\beta \epsilon_+^2 \right) \, ,
\end{align}
where $ \varphi_\alpha $ denotes the VEV of the scalar field in the tensor multiplet, $ \phi_a $ is the holonomy for the $ a $-th symmetry group including gauge and flavor symmetries, $ \langle \cdot, \cdot \rangle $ is the symmetric bilinear Killing form of the symmetry algebra, and $ \epsilon_+ = (\epsilon_1+\epsilon_2)/2 $.

The effective prepotential also contains the 1-loop contributions of the vector and hypermultiplets. Suppose that the low energy theory has a gauge algebra $ \mathfrak{g}=\oplus_\alpha \mathfrak{g}_\alpha $ and hypermultiplets in a representation $ \mathbf{R} $ of $ \mathfrak{g} $ which decomposes into irreducible representations as $ \mathbf{R} = \oplus_f \mathbf{R}_f $. Then the 1-loop contribution of these multiplets to the effective prepotential on the $ \Omega $-background is \cite{Kim:2020hhh, Intriligator:1997pq}
\begin{align}
    \begin{aligned}
        \epsilon_1 \epsilon_2 \mathcal{E}_0^{\text{1-loop}}
        &= \frac{1}{12} \sum_{n\in \mathbb{Z}} \sum_{\rho \in \Delta} \left|n\tau + \langle \rho, \phi \rangle\right|^3 - \frac{1}{12} \sum_{n\in \mathbb{Z}} \sum_f \sum_{\rho \in w(\mathbf{R}_f)} \left| n\tau + \langle \rho, \phi \rangle + m_f \right|^3 \\
        &\quad - \frac{\epsilon_1^2 + \epsilon_2^2}{48} \biggl( \sum_{\rho \in \Delta} |\langle \rho, \phi \rangle| - \sum_f \sum_{\rho \in w(\mathbf{R}_f)} |\langle \rho, \phi \rangle + m_f | \biggr)  + \frac{\epsilon_+^2}{4} \sum_{\rho \in \Delta} |\langle \rho, \phi \rangle| \, ,
    \end{aligned}
\end{align}
where $ \Delta $ is the set of roots of $ \mathfrak{g} $, and $ w(\mathbf{R}_f) $ and $ m_f $ are the weight system of $ \mathbf{R}_f $ and the mass parameter of the $ f $-th hypermultiplet, respectively. The absolute values are defined with respect to a chosen chamber of the parameters, and the summation over the KK-momentum $ n $ is understood with zeta-function regularization. The first line arises from the effective cubic Chern-Simons actions of the 5d KK theory, while the second line comes from the mixed gauge-gravitational and gauge-$ \SU{2}_R $ Chern-Simons actions.

We now consider the full partition function \eqref{eq:Z} on the $ \Omega $-background. It takes the factorized form
\begin{align}\label{eq:Z-fact}
    Z = e^{-2\pi i \mathcal{E}_0} \cdot Z_{\mathrm{pert}} \cdot \sum_{k} e^{-2\pi i \Omega^{\alpha\beta} k_\alpha \varphi_\beta} Z_{k} \, .
\end{align}
Here, $ \mathcal{E}_0 = \mathcal{E}_0^{\text{tree}} + \mathcal{E}_0^{\text{1-loop}} $ is the effective prepotential, $ Z_{\mathrm{pert}} = Z_{\mathrm{pert}}^{\mathrm{vec}} Z_{\mathrm{pert}}^{\mathrm{hyp}} $ is the 1-loop perturbative contribution of the vector and hypermultiplets given by
\begin{align}\label{eq:pert}
    \begin{aligned}
        Z_{\mathrm{pert}}^{\mathrm{vec}} &= \PE\biggl[ -\frac{1+p_1p_2}{(1-p_1)(1-p_2)(1-q)} \sum_{\rho \in \Delta^+} \Bigl( e^{2\pi i \langle \rho, \phi \rangle} + q e^{-2\pi i \langle \rho,\phi \rangle} \Bigr)  \biggr] \, , \\
        Z_{\mathrm{pert}}^{\mathrm{hyp}} &= \PE\biggl[ \frac{\sqrt{p_1p_2}}{(1-p_1)(1-p_2)(1-q)}  \sum_f \sum_{\rho \in w(\mathbf{R}_f)} \Bigl(e^{2\pi i | \langle \rho, \phi \rangle + m_f|} + q e^{-2\pi i | \langle \rho, \phi \rangle + m_f|} \Bigr) \biggr] \, ,
    \end{aligned}
\end{align}
where $ \PE[\cdot] $ is the plethystic exponential defined in Appendix~\ref{app:special}, we denote $ p_j=e^{2\pi i \epsilon_j} $ and $ q=e^{2\pi i \tau} $, and $ Z_k $ is the elliptic genus of the self-dual strings with tensor charge $ k=(k_\alpha) $ with $ Z_0=1 $. Note that although the partition function also contains the 1-loop contribution of the tensor multiplets, we omit it because it does not depend on the dynamical parameters $ \varphi $ and $ \phi $ of the theory.

If an explicit 2d gauge theory on the self-dual strings is known, one can compute the elliptic genus $ Z_k $ by supersymmetric localization. However, such a description is available only for a limited class of theories. The blowup equations \cite{Nakajima:2003pg, Nakajima:2005fg, Gottsche:2006bm} provide an alternative method for computing the elliptic genera. The blowup equations are functional equations satisfied by the full partition function \eqref{eq:Z-fact}, which relate the partition function on $ \mathbb{C}^2 \times T^2 $ to that on its $ \hat{\mathbb{C}}^2 \times T^2 $, which is obtained by replacing the origin of $ \mathbb{C}^2 $ with the exceptional 2-cycle $ \mathbb{P}^1 $. They take the form 
\begin{align}\label{eq:blowup}
    \Lambda(m, \epsilon_1, \epsilon_2) \hat{Z}(\varphi, \phi,m, \tau; \epsilon_1, \epsilon_2)
    = \sum_{\ell, n} (-1)^{|\ell|+|n|} \hat{Z}^{(N)}(\ell,n, B) \hat{Z}^{(S)}(\ell,n, B) \, ,
\end{align}
where $ \Lambda $ is a factor independent of the dynamical parameters $ \varphi $ and $ \phi $ of the theory. The hatted partition function $ \hat{Z} $ differs from the original partition function $ Z $ in \eqref{eq:Z} by the replacement $ (-1)^F \to (-1)^{2J_R} $. This is equivalent to the shift $ \epsilon_1 \to \epsilon_1+1 $ in all factors of \eqref{eq:Z-fact} except for the prepotential $ \mathcal{E}_0 $ contribution.\footnote{\label{footnote:shift}In practice, one may instead shift only the classical factor $ e^{-2\pi i \mathcal{E}_0} $ in \eqref{eq:Z-fact} and keep the other factors unchanged. In the following sections, where we write the blowup equations explicitly for a given 6d theory, we use this convention, since it displays the identities for the elliptic genera more clearly.} The two factors $ \hat{Z}^{(N)} $ and $ \hat{Z}^{(S)} $ are defined as
\begin{align}
    \begin{aligned}\label{eq:NSpole-Z}
        \hat{Z}^{(N)}(\ell,n,B) &= \hat{Z}(\varphi+\ell\epsilon_1, \phi+n\epsilon_1, m+B\epsilon_1, \tau; \epsilon_1, \epsilon_2-\epsilon_1) \, , \\
        \hat{Z}^{(S)}(\ell,n,B) &= \hat{Z}(\varphi+\ell\epsilon_2, \phi+n\epsilon_2, m+B\epsilon_2, \tau; \epsilon_1-\epsilon_2, \epsilon_2) \, ,
    \end{aligned}
\end{align}
where $ \ell $ and $ n $ are the dynamical magnetic fluxes for the tensor and gauge symmetries, respectively, and $ B $ denotes the background magnetic fluxes for the flavor symmetry. These fluxes are turned on along the exceptional $ \mathbb{P}^1 $, and $ |\ell|=\sum \ell_\alpha $ and $ |n|=\sum n_a $. These two factors are interpreted as the partition functions localized at the north and south poles of $ \mathbb{P}^1 $. For the blowup equations to be consistent, the magnetic fluxes must satisfy a quantization condition: for every BPS state with tensor charge $ k $, gauge charge $ e $, flavor charge $ f $ and spins $ (j_l, j_r) $ under $ \SU{2}_l \times \SU{2}_r $ rotation, the magnetic fluxes satisfy \cite{Huang:2017mis, Kim:2020hhh}
\begin{align}
    -\Omega^{\alpha\beta} k_\alpha \ell_{\beta} + \langle e, n \rangle + f \cdot B \in \mathbb{Z} + j_l + j_r + \frac{1}{2} \, .
\end{align}
For the vector multiplets which have $ (j_l,j_r)=(0,\frac{1}{2}) $, this requires $ \langle \rho, n \rangle \in \mathbb{Z} $ for every root $ \rho $, while the hypermultiplets with $ (j_l,j_r)=(0,0) $ require $ \langle \rho, n \rangle + B_f \in \mathbb{Z}+\frac{1}{2} $ for every weight vector $ \rho \in w(\mathbf{R}_f) $. Similarly, the self-dual strings fix the fractional parts of the tensor fluxes $ \ell $. Given the effective prepotential $ \mathcal{E}_0 $, the perturbative partition function \eqref{eq:pert}, and a consistent set of magnetic fluxes, one can solve the blowup equation order by order in the fugacities in the theory to compute the elliptic genera. A more detailed explanation for the blowup equations can be found in \cite{Kim:2020hhh}.

\subsection{Codimension-4 defects} \label{sec:codim4}

The 6d SCFTs admit half-BPS defect operators of codimension four. When the 6d theory is placed on $ \mathbb{C}^2 \times T^2 $, such defects wrap $ T^2 $ and are localized at the origin of $ \mathbb{C}^2 $. In this subsection, we consider two realizations of these defects, by 2d chiral fermions coupled to the bulk gauge field and by Wilson surfaces. We also discuss their partition functions on the $ \Omega $-background and the relation between them.

We first consider the codimension-4 defect realized by a 2d chiral fermion $ \psi $ localized at the origin of $ \mathbb{C}^2 $ and coupled to the bulk gauge symmetry in a representation $ \mathbf{R} $ of the gauge algebra. This defect is also known as the 6d qq-character \cite{Nekrasov:2015wsu, Kim:2016qqs}. Concretely, we add the 2d action
\begin{align}
    S_{\mathrm{2d}} = \int d^2x \, \bar{\psi}_- (D_0 + D_1) \psi_- \, ,
\end{align}
to the 6d action, where $ D_j = \partial_j + iA_j $ for the 6d bulk gauge field $ A_j $. This action introduces a codimension-4 defect that preserves half of the supersymmetry. However, the chiral fermion has a gauge anomaly
\begin{align}\label{eq:I4-codim4}
    I_4^{\text{codim-4}} \supset \frac{1}{2} \tr_{\mathbf{R}} F_a^2 = \frac{1}{2} A_{a\mathbf{R}} \Tr F_a^2 \, ,
\end{align}
where $ I_4^{\text{codim-4}} $ is the anomaly 4-form of the 2d fermions, and $ A_{a\mathbf{R}} $ is the quadratic index of $ \mathbf{R} $ with respect to the $ a $-th gauge algebra, defined by $ \tr_{\mathbf{R}} F_a^2 = A_{a\mathbf{R}} \Tr F_a^2 $. Its values for the simple Lie algebras can be found in Appendix~\ref{app:Lie}. There is no mixed gauge-flavor anomaly term for a semisimple gauge algebra, since $ \tr_{\mathbf{R}} F = 0 $, and we omit the 't Hooft anomaly terms involving only the background fields in the anomaly 4-form. This gauge anomaly must be canceled, which is achieved by adding the Green-Schwarz counterterm
\begin{align}\label{eq:codim4-ct}
    S_{\mathrm{ct}} = 2\pi i \sum_\alpha \kappa^\alpha \int_\Sigma B_\alpha \, , \qquad
    I_4^{\text{inflow}} = -\sum_\alpha \kappa^\alpha X_{4\alpha} \, ,
\end{align}
where $ \kappa^\alpha $ are constants, and $ I_4^{\text{inflow}} $ is the contribution of the counterterm to the anomaly 4-form. Requiring that the inflow cancels the gauge anomaly gives
\begin{align}\label{eq:anomaly-codim4}
    \sum_a \left( \frac{1}{2} A_{a\mathbf{R}} - \frac{1}{4} \sum_\alpha \kappa^\alpha b_{a \alpha} \right) \Tr F_a^2 = 0 \quad \Rightarrow \quad \sum_\alpha \kappa^\alpha b_{a \alpha} = 2A_{a\mathbf{R}} \, ,
\end{align}
for every $ a $ labeling a gauge symmetry. This constrains the tensor charge $ \kappa^\alpha $ of the probe string.

The counterterm \eqref{eq:codim4-ct} contributes to the classical part of the partition function as
\begin{align}
    \mathcal{E}_1 = \sum_\alpha \kappa^\alpha \varphi_\alpha \, ,
\end{align}
which is linear in the tensor parameters $ \varphi_\alpha $ and independent of $ \epsilon_1 $ and $ \epsilon_2 $, since the defect sits at the fixed point of the $ \Omega $-background. The 1-loop contribution of the 2d fermion to the partition function is
\begin{align}\label{eq:X0}
    \mathcal{X}_0 = \prod_{\rho \in w(\mathbf{R})} \frac{\theta_1(\tau, x+\langle \rho, \phi \rangle)}{\eta(\tau)} \, ,
\end{align}
where $ x $ is the holonomy along $ T^2 $ for the $ \mathrm{U}(1) $ flavor symmetry acting on the fermion. The full partition function is expressed as
\begin{align}\label{eq:codim4-X}
    \mathcal{X} = e^{-2\pi i (\mathcal{E}_0 + \mathcal{E}_1)} \cdot Z_{\mathrm{pert}} \cdot \sum_k e^{-2\pi i \Omega^{\alpha \beta} k_\alpha \varphi_\beta} \mathcal{X}_k \, ,
\end{align}
where $ \mathcal{X}_k $ is the elliptic genus of the self-dual strings with tensor charge $ k=(k_\alpha) $ in the presence of the codimension-4 defect. The defect partition function $ \mathcal{X} $ also satisfies blowup equations \cite{Kim:2021gyj}. As in \eqref{eq:codim4-X}, one needs to include the contribution of $ \mathcal{E}_1 $, which provides an overall factor of the tensor fugacity and has a non-trivial effect in the blowup equation. One can use the same magnetic fluxes $ (\ell, n, B) $ as in the blowup equation \eqref{eq:blowup}, together with the zero magnetic flux $ B_x=0 $ for the defect flavor symmetry. Then the blowup equation is given by \cite{Kim:2021gyj, Wang:2023zcb}
\begin{align}\label{eq:blowup-codim4}
    \Lambda_{\mathcal{X}}(m,\epsilon_1,\epsilon_2) \hat{\mathcal{X}}
    = \sum_{\ell,n} (-1)^{|\ell|+|n|} \hat{\mathcal{X}}^{(N)}(\ell,n,B) \hat{Z}^{(S)}(\ell,n,B) \, ,
\end{align}
where $ \Lambda_{\mathcal{X}} $ is again independent of the dynamical parameters, and $ \hat{\mathcal{X}}^{(N)} $ is the north-pole defect partition function defined in the same way as in \eqref{eq:NSpole-Z}.\footnote{When the representation $ \mathbf{R} $ of the codimension-4 defect is large, the blowup equation can have ambiguities~\cite{Kim:2021gyj}. To avoid this difficulty, we assume that $ \mathbf{R} $ is a minuscule or quasi-minuscule representation.} The structure of the equation suggests that the codimension-4 defect is inserted at the north pole of the blown-up geometry $ \hat{\mathbb{C}}^2 \times T^2 $, which becomes the original defect localized at the origin of $ \mathbb{C}^2 \times T^2 $ after the blowdown. See \cite{Kim:2021gyj} for more details.

We now turn to the Wilson surfaces, which are the 6d analogues of Wilson loop operators. On the tensor branch, a Wilson surface carrying tensor charge $\kappa^\alpha$ and associated with a representation $\mathbf{R}$ of the gauge algebra is formally defined in terms of the integral of the two-form potential $B$ over the worldsheet $\Sigma$ of a probe string \cite{Ganor:1996nf,Chen:2007ir,Agarwal:2018tso,Apruzzi:2020zot}:
\begin{align}
    \mathcal{W}_{\mathbf{R}}[\Sigma] = \Tr_{\mathbf{R}} \mathcal{P} \exp\left(i\kappa^\alpha\int_{\Sigma} B_\alpha + \cdots\right) \, ,
\end{align}
where $ \cdots $ denotes the supersymmetric completion. Upon compactification of the 6d theory on a circle, a Wilson surface reduces to a Wilson loop operator in the 5d KK theory. These Wilson loops can be defined in terms of the geometry of the local Calabi-Yau threefold that engineers the 5d KK theory in M-theory compactification \cite{Kim:2021gyj}. In this geometric realization, a Wilson loop operator is represented by an M2-brane wrapping a primitive non-compact holomorphic curve $ \mathcal{C} $, while M2-branes wrapping $ \mathcal{C} + \sum_s n_s C_s $ with effective compact curves $ C_s $ and $ n_s \geq 0 $ describe the BPS states bound to the probe. Here, the non-compact curve $ \mathcal{C} $ is called primitive if it is not the sum of another non-compact curve and effective compact curves $ C_s $. It is characterized by its intersection numbers with the compact divisors, which determine the electric charges of the Wilson loop operator.

This motivates the following field-theoretic definition of the fundamental Wilson surfaces. Suppose that the 6d theory has a gauge algebra $ \mathfrak{g}=\oplus_\alpha \mathfrak{g}_\alpha $, where each $ \mathfrak{g}_\alpha $ is supported on the $ \alpha $-th tensor node. For each $ \alpha $, let $ \omega_i $ be the fundamental weights of $ \mathfrak{g}_\alpha $ and $ d_i^\vee $ the corresponding dual Coxeter labels. We also denote $ \hat{\omega}_i $ the fundamental weights of the affine Lie algebra $ \hat{\mathfrak{g}}_\alpha $. The convention for the Lie algebra can be found in Appendix~\ref{app:Lie}. We then define the fundamental Wilson surface $ \mathcal{W}_{\alpha i} $ as the Wilson surface operator whose VEV at the perturbative level is given by the character of the integrable representation with highest weight $ \hat{\omega}_{i} $ of $ \hat{\mathfrak{g}}_\alpha $ at level $ d_i^\vee $. We write the Wilson surface VEV $ W_{\alpha i} $ as the normalized partition function $ \mathcal{W}_{\alpha i}/Z $ in the presence of the Wilson surface and expand it in the string fugacities as
\begin{align}\label{eq:wilsonVEV}
    W_{\alpha i} = \frac{\mathcal{W}_{\alpha i}}{Z} = e^{-2\pi i d_i^\vee \varphi_\alpha} \sum_{k} e^{-2\pi i \Omega^{\beta\gamma} k_\beta \varphi_\gamma} W_{\alpha i}^{(k)} \, ,
\end{align}
where we factor out the overall $ e^{-2\pi i d_i^\vee \varphi_\alpha} $ factor, which accounts for the level of the affine representation. The definition of the fundamental Wilson surface then means that its perturbative part is given by
\begin{align}\label{eq:W-affine}
    W_{\alpha i}^{(0)} = \chi_{\hat{\omega}_{i}}^{\hat{\mathfrak{g}}_\alpha} = q^{\Delta_i} \left( \chi_{\omega_{i}}^{\mathfrak{g}_\alpha}(\phi) + \mathcal{O}(q) \right) \, .
\end{align}
Here, $ \chi_{\hat{\omega}_{i}}^{\hat{\mathfrak{g}}_\alpha} $ is the level $ d_i^\vee $ affine character of $ \hat{\mathfrak{g}}_\alpha $ for the highest weight $ \hat{\omega}_{i} $, $ \chi_{\omega_{i}}^{\mathfrak{g}_\alpha} $ is the character of the highest weight $ \omega_{i} $ representation of $ \mathfrak{g}_\alpha $, and the leading order exponent $ \Delta_i $ of $ q $ is given by
\begin{align}
    \Delta_i = \frac{\langle \omega_{i}, \omega_{i}+2\rho \rangle}{2(d_i^\vee + h^\vee)} - \frac{d_i^\vee \dim \mathfrak{g}_\alpha}{24(d_i^\vee + h^\vee)} \, ,
\end{align}
where $ \rho=\sum_i \omega_i $ is the Weyl vector and $ h^\vee $ is the dual Coxeter number of $ \mathfrak{g}_\alpha $. At leading order in $ q $, $ W_{\alpha i} $ contains the state with tensor charge $ d_i^\vee $ and gauge weight $ -\omega_{i^*} = w_0 \omega_i $, where $ w_0 $ is the longest Weyl group element, so that $ -\omega_{i^*} $ is the lowest weight of the finite-dimensional highest weight representation of $ \mathfrak{g}_\alpha $ with highest weight $ \omega_i $. This term contributes $ e^{-2\pi i d_i^\vee \varphi_\alpha} e^{-2\pi i \langle \omega_{i^*}, \phi \rangle} $. This is the state of the string probe itself. In the geometric realization, it corresponds to the primitive non-compact curve $ \mathcal{C} $ intersecting the compact divisor associated with the K\"ahler parameter $ \langle \omega_{i^*}, \phi \rangle $ once and not intersecting with the other compact divisors. We expect that the full fundamental Wilson surface partition function $ \mathcal{W}_{\alpha i} $ again satisfies the blowup equations \eqref{eq:blowup-codim4} with $ \mathcal{X} $ replaced by $ \mathcal{W}_{\alpha i} $, and the higher string corrections $ W_{\alpha i}^{(k)} $ of the Wilson surface expectation value can be computed from them.

As the 5d qq-character is a generating function of Wilson loops \cite{Nekrasov:2015wsu, Kim:2016qqs}, it is possible to decompose the codimension-4 defect partition function $ \mathcal{X} $ in terms of the partition functions of Wilson surfaces. The 1-loop contribution \eqref{eq:X0} is a holomorphic function of $ \phi $ invariant under the Weyl group of the gauge algebra, and is also quasi-periodic under the shift $ \phi \to \phi + \tau \beta^\vee $,
\begin{align}
    \mathcal{X}_0(\phi + \tau \beta^\vee) = q^{-\frac{K_\alpha}{2} \langle \beta^\vee, \beta^\vee \rangle} e^{-2\pi i K_\alpha \langle \beta^\vee, \phi \rangle} \mathcal{X}_0(\phi) \, , \quad
    K_\alpha = 2A_{\alpha \mathbf{R}} \, ,
\end{align}
for a coroot $ \beta^\vee $ of $ \mathfrak{g}_\alpha $. This is precisely the transformation property of the characters of $ \hat{\mathfrak{g}}_\alpha $ at level $ K_\alpha $, and the characters of the integrable highest weight representations at this level form a basis of the space of such Weyl-invariant holomorphic functions \cite{Kac:1990gs}. The level of a product is the sum of levels of its factors. This space is also spanned by the monomials in the fundamental characters \eqref{eq:W-affine} of total level $ K_\alpha $. Hence, $ \mathcal{X}_0 $ can be decomposed in terms of the perturbative VEVs of the fundamental Wilson surfaces as
\begin{align}\label{eq:X0-decomp}
    \mathcal{X}_0 = \sum_{N} C_N(\tau, x) \prod_\alpha \prod_{i=0}^{\rk \mathfrak{g}_\alpha} \bigl( W_{\alpha i}^{(0)} \bigr)^{N_{\alpha i}} \, , \quad
    \sum_i d_i^\vee N_{\alpha i}=K_\alpha \, ,
\end{align}
where $ N=(N_{\alpha i}) $ runs over the non-negative integers satisfying the level constraint written in the second equation, and the coefficients $ C_N $ depend only on $ \tau $ and $ x $ but not on the dynamical parameters $ \varphi $ and $ \phi $, while $ W_{\alpha i}^{(0)} $ does not depend on the defect parameter $ x $. The prepotential factor $ \mathcal{E}_1 $ of $ \mathcal{X} $ is also reproduced by the overall $ \varphi $ factors in \eqref{eq:wilsonVEV}. If this decomposition is linear in $ W_{\alpha i}^{(0)} $, as happens when only a single gauge factor has level one but all other levels vanish 
\begin{align}
 K_{\alpha_0}=1\quad \mathrm{and}\quad K_\alpha=0\quad (\alpha\neq \alpha_0) \ ,   
\end{align}
then the decomposition \eqref{eq:X0-decomp} extends to all orders in the string fugacities as
\begin{align}
    \frac{\mathcal{X}}{Z} = \sum_{N} C_N(\tau, x) \prod_\alpha \prod_{i=0}^{\rk \mathfrak{g}_\alpha} \bigl( W_{\alpha i} \bigr)^{N_{\alpha i}} \, ,
\end{align}
since the blowup equation is linear in the Wilson surface partition function and the coefficients $ C_N(\tau,x) $ are not shifted by the magnetic fluxes. On the other hand, if \eqref{eq:X0-decomp} is not linear in the Wilson surface VEVs, we cannot say that the decomposition extends to higher string order. Nevertheless for the codimension-4 defects realized by 2d fermions in fundamental representations considered in later sections, we find that the decomposition extends to higher string orders in the NS-limit $ \epsilon_2 \to 0 $ \cite{Huang:2022hdo}.

\subsection{Codimension-2 defects} \label{sec:codim2}

We now turn to the codimension-2 defects realized by 4d $ \mathcal{N}=1 $ chiral multiplets supported on $ \mathbb{C}_{\epsilon_1} \times T^2 $ and localized at the origin of $ \mathbb{C}_{\epsilon_2} $, coupled to the bulk gauge fields. As in the codimension-4 case, the defect matter contributes to the gauge anomaly. Suppose a 4d chiral multiplet is coupled to the bulk gauge symmetry in a representation $ \mathbf{R}_I $, and carries charge $ u_I $ under the $ \U{1}_x $ flavor symmetry. Then the defect matter contributes to the anomaly 6-form as
\begin{align}
    I_6^{\text{codim-2}} \supset \frac{1}{6} \tr_{\mathbf{R}_I} F_a^3 + \frac{1}{2} u_I F_x \tr_{\mathbf{R}_I} F_a^2 \, ,
\end{align}
where $ F_x $ is the background 2-form field strength of $ \U{1}_x $, and we omit the 't Hooft anomaly terms involving only background flavor fields. The first term is the cubic gauge anomaly. It cannot be canceled by the anomaly inflow from the counterterm involving the 2-form fields $ B_\alpha $, since the inflow from the Green-Schwarz term has the form $ X_2 \wedge X_{4\alpha} $ with a 2-form $ X_2 $. We therefore require that this term cancel among the defect multiplets, $ \sum_I \tr_{\mathbf{R}_I} F_a^3 = 0 $.\footnote{However, this condition assumes that there is no additional contribution from the bulk fermions. In theories with 6d charged hypermultiplets, a suitable flavor background can induce a localized contribution from their mixed gauge-flavor anomaly. This may cancel the cubic gauge anomaly of the defect matter. Examples are the defects considered in \cite{Chen:2020jla}.} This implies that, for instance, a codimension-2 defect with a single 4d chiral multiplet in the fundamental representation $ \mathbf{3} $ of $ \su 3 $ is not allowed in the pure $ \su 3 $ gauge theory, while $ \mathbf{3}\oplus \bar{\mathbf{3}} $ is allowed. The second term is the mixed gauge-$ \U{1}_x $ anomaly, which can be canceled by anomaly inflow. We add the counterterm $ S_{\mathrm{ct}} $ to the action, which gives the inflow $ I_6^{\text{inflow}} $ as
\begin{align}
    S_{\mathrm{ct}} = 2\pi i \sum_\alpha \kappa^\alpha \int B_\alpha \wedge F_x \, , \quad
    I_6^{\text{inflow}} = -\sum_\alpha \kappa^\alpha F_x \wedge X_{4\alpha} \, ,
\end{align}
where $ X_{4\alpha} $ is the 4-form given in \eqref{eq:X4}. The cancellation of the mixed anomaly requires
\begin{align}\label{eq:codim2-anomaly}
    \sum_a \left( \sum_I \frac{1}{2} u_I A_{a\mathbf{R}_I} - \frac{1}{4} \sum_\alpha \kappa^\alpha b_{a\alpha} \right) \Tr F_a^2 = 0 \quad \Rightarrow \quad \sum_\alpha \kappa^\alpha b_{a\alpha} = 2\sum_I u_I A_{a\mathbf{R}_I} \, ,
\end{align}
where $ A_{a\mathbf{R}} $ is the quadratic index of the representation $ \mathbf{R} $. This counterterm contributes to the effective prepotential as
\begin{align}
    \mathcal{E}_2^{\text{tree}}
    = \frac{x}{\epsilon_1} \sum_\alpha \kappa^\alpha \varphi_\alpha \, ,
\end{align}
where $ 1/\epsilon_1 $ is the equivariant volume of $ \mathbb{C}_{\epsilon_1} $.

We now consider the full partition function in the presence of a codimension-2 defect. The 1-loop contribution of the 4d chiral multiplet is given by
\begin{align}\label{eq:psi-pert1}
    \Psi_{\mathrm{pert}}
    &= \PE\left[ \frac{\sqrt{p_1}}{(1-p_1)(1-q)} \sum_I \sum_{\rho \in w(\mathbf{R}_I)} \left( e^{2\pi i (u_I x + \langle \rho, \phi \rangle)} - qe^{-2\pi i (u_I x + \langle \rho, \phi \rangle)} \right) \right] \nonumber \\
    &= \prod_I \prod_{\rho \in w(\mathbf{R}_I)} \Gamma_e\Bigl(e^{2\pi i (u_I x + \langle \rho, \phi \rangle +\epsilon_1/2)}; p_1, q \Bigr) \, ,
\end{align}
where $ x $ is the holonomy of the $ \U{1}_x $ symmetry, and $ \Gamma_e $ is the elliptic gamma function defined in Appendix~\ref{app:special}. The effective prepotential also receives 1-loop contributions from the defect matter. The circle reduction of the 4d chiral multiplets gives a tower of 3d $ \mathcal{N}=2 $ chiral multiplets, and the effective Chern-Simons coupling contributes to the effective prepotential as \cite{Closset:2017zgf, Corvilain:2017luj}
\begin{align}
    \mathcal{E}_2^{\text{1-loop}}
    = \frac{1}{4\epsilon_1} \sum_{n\in \mathbb{Z}} \sum_I \sum_{\rho \in w(\mathbf{R}_I)} \left( n\tau + \langle \rho, \phi \rangle + u_I x \right)^2 \sgn\left( n\tau + \langle \rho, \phi \rangle + u_I x \right) \, ,
\end{align}
where $ \sgn $ is the sign function. Summing over the KK modes in the region $ \tau \gg x, \phi $, and using the zeta-function regularization and $ \sum_{\rho \in w(\mathbf{R}_I)} \rho = 0 $ for the semisimple gauge algebra, this becomes
\begin{align}
    \mathcal{E}_2^{\text{1-loop}} = \frac{1}{4\epsilon_1} \sum_I \sum_{\rho \in w(\mathbf{R}_I)} \sgn\bigl( u_I x + \langle \rho, \phi \rangle \bigr) \bigl( u_I x + \langle \rho, \phi \rangle  \bigr)^2 \, ,
\end{align}
up to terms independent of $ \phi $. The defect contribution to the effective prepotential is the sum of $ \mathcal{E}_2^{\text{tree}} $ and $ \mathcal{E}_2^{\text{1-loop}} $:
\begin{align}
    \mathcal{E}_2 = \mathcal{E}_2^{\text{tree}} + \mathcal{E}_2^{\text{1-loop}} \, ,
\end{align}
and the full codimension-2 defect partition function is expressed as
\begin{align}
    \Psi = e^{-2\pi i (\mathcal{E}_0 + \mathcal{E}_2)} \cdot Z_{\mathrm{pert}} \cdot \Psi_{\mathrm{pert}} \cdot \sum_k e^{-2\pi i \Omega^{\alpha\beta} k_\alpha \varphi_\beta} \Psi_k \, ,
\end{align}
where $ \Psi_0=1 $ and $ \Psi_k $ is the elliptic genus of the self-dual strings with tensor charge $ k $ in the presence of the defect. We expect that this defect partition function is also a solution of the blowup equations of the form \cite{Kim:2025qaf}
\begin{align}
    \Lambda_\Psi(m,x,\tau; \epsilon_1,\epsilon_2) \hat{\Psi} = \sum_{\ell,n} (-1)^{|\ell|+|n|} \hat{\Psi}^{(N)} \hat{Z}^{(S)} \, ,
\end{align}
where $ \hat{\Psi}^{(N)} $ is defined analogously to \eqref{eq:NSpole-Z} for $ \hat{\Psi} $. In this paper, we turn on the same magnetic fluxes as in the original blowup equation \eqref{eq:blowup} and set $ B_x=0 $, although it is also possible to turn on a non-zero flux for $ x $. The structure of the blowup equation reflects the fact that the codimension-2 defect is inserted along $ \mathbb{C}_{\epsilon_1} $. On the blown-up geometry $ \hat{\mathbb{C}}^2 $, the defect meets the exceptional $ \mathbb{P}^1 $ only at the north pole. Hence the defect contributes only to the north pole factor, while the south pole factor is the partition function $ \hat{Z}^{(S)} $ without a defect. See \cite{Kim:2025qaf} for details and more examples.

Before we move to the discussion of the theories on a $ -3 $ curve in the next section, we comment on a different but equivalent representation of the 1-loop contribution of the codimension-2 defect. For simplicity, we focus on the single tensor node case and assume that all the defect multiplets carry the $ \U{1}_x $ charge $ +1 $. Let $ \mathbf{R} $ be their total representation of the bulk gauge algebra and assume that its weight system is symmetric, $ w(\mathbf{R}) = -w(\mathbf{R}) $. In this case, the cubic gauge anomaly cancellation $ \tr_{\mathbf{R}} F^3=0 $ is automatic. Then we write the 1-loop perturbative partition function as
\begin{align}\label{eq:psi-pert2}
    \Psi_{\mathrm{pert}}' &= \PE\left[ \frac{\sqrt{p_1} (e^{2\pi i x}-e^{-2\pi i x})}{(1-p_1)(1-q)} \sum_{\rho\in \mathcal{S}} \left( e^{2\pi i \langle \rho, \phi \rangle} + q e^{-2\pi i \langle \rho, \phi \rangle} \right) \right] \nonumber \\
    &= \prod_{\rho\in \mathcal{S}} \Gamma_e\bigl(e^{2\pi i (x+\langle \rho, \phi \rangle+\epsilon_1/2)}; p_1,q\bigr) \Gamma_e\bigl(e^{2\pi i (x-\langle \rho, \phi \rangle+\epsilon_1/2 + \tau)}; p_1,q\bigr) \, ,
\end{align}
where $ \mathcal{S} \cup (-\mathcal{S}) = w(\mathbf{R})\setminus \{0\} $ divides the non-zero weight vectors into two sets. Zero weights give a $ \phi $-independent factor, which we omit. Compared with \eqref{eq:psi-pert1}, the only difference is at order $ q^0 $ inside the plethystic exponential:
\begin{align}
    \frac{\Psi_{\mathrm{pert}}'}{\Psi_{\mathrm{pert}}} = \PE\left[ -\frac{\sqrt{p_1}}{1-p_1} \sum_{\rho \in \mathcal{S}} \left( e^{2\pi i (x-\langle \rho, \phi \rangle)} + e^{-2\pi i (x - \langle \rho, \phi \rangle)} \right) \right] = \prod_{\rho \in \mathcal{S}} \vartheta(e^{2\pi i (x - \langle \rho, \phi \rangle + \epsilon_1/2)}; p_1) \, ,
\end{align}
where $ \vartheta $ is the $ q $-theta function defined in Appendix~\ref{app:special}. This corresponds to changing the boundary condition of the 4d chiral multiplets at infinity in $ \mathbb{C}_{\epsilon_1} $ \cite{Longhi:2019hdh}. In this representation, the 1-loop contribution to the effective prepotential becomes
\begin{align}\label{eq:codim2-pre-loop}
    \mathcal{E}_2^{\text{1-loop}'} = \frac{1}{4\epsilon_1} \sum_{\rho \in \mathcal{S}} \left( (x+\langle \rho, \phi \rangle)^2 - (x-\langle \rho, \phi \rangle)^2 \right) = \frac{x}{\epsilon_1} \sum_{\rho \in \mathcal{S}} \langle \rho, \phi \rangle \, .
\end{align}
The two representations of the 1-loop contribution are related by analytic continuation, and the choice does not affect the elliptic genera $ \Psi_k $.

\section{6d pure \texorpdfstring{$ \su 3 $}{su(3)} gauge theory} \label{sec:su3}

In this section, we study in detail the 6d SCFT whose tensor branch is described by the pure $ \su 3 $ gauge theory. In F-theory, it is engineered on an elliptically fibered Calabi-Yau threefold whose base contains a $ -3 $ curve, over which the elliptic fiber is of Kodaira type IV. This theory is a non-Higgsable cluster of 6d SCFTs with $ \Omega^{\alpha\beta}=(-3) $. We first review the partition function without defects and the blowup equations that determine it, and then consider codimension-4 and codimension-2 defects and the quantum Seiberg--Witten geometry of this theory.

\subsection{Partition function and blowup equations}

We begin with the partition function without defects. The effective prepotential consists of the tree-level part 
\begin{align}\label{eq:su3-tree}
    \epsilon_1 \epsilon_2 \mathcal{E}_0^{\mathrm{tree}}
    &= \frac{3}{2}\tau \varphi^2 + 3\varphi \left( -\frac{\epsilon_1^2 + \epsilon_2^2}{12} + \left(\phi_1^2 - \phi_1 \phi_2 + \phi_2^2 \right) + \epsilon_+^2 \right) \, ,
\end{align}
and the 1-loop contribution of the vector multiplet
\begin{align}
    \begin{aligned}
        \epsilon_1 \epsilon_2 \mathcal{E}_0^{\text{1-loop}}
        &= \frac{1}{6} \left( 8\phi_1^3 - 3\phi_1^2\phi_2 - 3\phi_1 \phi_2^2 + 8\phi_2^3 \right) - \frac{\tau}{2} \left( \phi_1^2 - \phi_1 \phi_2 + \phi_2^2 \right) \\
        &\quad - \frac{\epsilon_1^2 + \epsilon_2^2}{12} (\phi_1+\phi_2) + \epsilon_+^2 (\phi_1+\phi_2) \, ,
    \end{aligned}
\end{align}
where $ \varphi $ is the tensor branch parameter, and $ \phi=\phi_1 \alpha_1^\vee + \phi_2 \alpha_2^\vee $ denotes the $ \mathfrak{su}(3) $ gauge holonomy in the basis of simple coroots $ \alpha_i^\vee $ of the $ \mathfrak{su}(3) $ gauge algebra. The full partition function takes the form
\begin{align}\label{eq:su3-fullZ}
    Z = e^{-2\pi i \mathcal{E}_0} \cdot Z_{\mathrm{pert}} \cdot \left( 1 + \sum_{k=1}^\infty Q^{3k} Z_k \right) \, ,
\end{align}
where $ Q=e^{2\pi i \varphi} $, $ Z_{\mathrm{pert}} $ is the 1-loop perturbative partition function \eqref{eq:pert} of the vector multiplet and $ Z_k $ is the elliptic genus of $ k $ self-dual strings. The elliptic genera are determined by the blowup equations \cite{Gu:2018gmy, Kim:2020hhh}, which are labeled by the set of magnetic fluxes
\begin{align}\label{eq:su3-flux}
    \ell \in \mathbb{Z} + \nu \, , \ (\nu = \pm 1/6, 1/2) \, , \quad
    n_i \in \mathbb{Z} \, , \quad
    B_\tau = 0 \, ,
\end{align}
where $ \ell $ and $ n_{1,2} $ are the dynamical magnetic fluxes associated with $ \varphi $ and $ \phi_{1,2} $, respectively, and $ B_\tau $ is the background flux for $ \tau $. We write $ n=n_1 \alpha_1^\vee + n_2 \alpha_2^\vee $ for the flux associated to $ \su 3 $ gauge symmetry. After summing over $ \ell $, the blowup equation for each $ r $ at order $ Q^{3k} $ reads
\begin{align}
\thtftn{-\nu}{1/2} (3\tau, 2\epsilon_+) Z_k = \sum_{P_k} (-1)^{n_1+n_2} \thtftn{-\nu}{1/2}(3\tau, 2\epsilon_+ - \mu) \mathcal{A}(n) Z_{k_1}^{(N)} Z_{k_2}^{(S)} \, ,
\end{align}
where $ Z_0=1 $, $ Z_k^{(N)} $ and $ Z_k^{(S)} $ denote the $ k $-string elliptic genus at the north and south poles of the exceptional $ \mathbb{P}^1 $ as in \eqref{eq:NSpole-Z}, respectively, and we use the notation of footnote~\ref{footnote:shift}. Here, $ \mu $ is defined as
\begin{align}\label{eq:su3-mu}
    \mu = 3k_1 \epsilon_1 + 3k_2 \epsilon_2 + 3\langle n, n \rangle \epsilon_+ + 3\langle n, \phi \rangle \, ,
\end{align}
and the sum runs over the set $ P_k $ defined as
\begin{align}\label{eq:su3-P}
    P_k = \Big\{ (n_1, n_2, k_1, k_2) \,\Big|\, \frac{1}{2}\langle n, n \rangle + k_1 + k_2 = k \, , \ n_i \in \mathbb{Z} \, , \ k_1, k_2 \in \mathbb{N} \Big\} \, .
\end{align}
The theta function with characteristics $ \thtftn{-\nu}{1/2} $ is defined in Appendix~\ref{app:special} and arises from the $ \ell $ sum in the prepotential contribution. The coefficient $ \thtftn{-\nu}{1/2}(3\tau,2\epsilon_+) $ on the LHS plays the role of $ \Lambda $ in \eqref{eq:blowup}, and is independent of the dynamical parameters. The factor
\begin{align}\label{eq:su3-A0}
    \mathcal{A}(n) &= \prod_{\rho \in \Delta^+} \prod_{j_1=0}^{|\langle \rho, n \rangle|-1} \prod_{\delta=0}^1 \prod_{j_2=0}^{j_1-\delta} \frac{\eta(\tau)}{\theta_1(\tau, \langle \rho, \phi \rangle + s_\rho((j_2+\delta) \epsilon_1 + (j_1-j_2) \epsilon_2))} \, ,
\end{align}
with $ s_\rho = \sgn\langle \rho, n \rangle $ comes from the 1-loop contributions of the vector multiplet.

At the $ k $-th order, the unknown $ k $-string elliptic genus appears through $ Z_k $, $ Z_k^{(N)} $ and $ Z_k^{(S)} $, while all other terms involve only $ Z_{k'} $ with $ k'<k $. Treating these three functions as independent unknowns, the three blowup equations labeled by $ \nu=\pm 1/6, 1/2 $ form a linear system for the $ k $-string elliptic genus. However, for $ k=1 $, the determinant of the linear system vanishes, and the system cannot be directly inverted. One can instead determine the 1-string elliptic genus by combining a Gopakumar-Vafa expansion of the partition function with the blowup equations \cite{Kim:2020hhh} and the modular bootstrap method \cite{DelZotto:2017mee}. Alternatively, the ADHM-like description of the self-dual string \cite{Kim:2018gjo} gives the compact expression
\begin{align}\label{eq:su3-Z1}
    Z_1 = \sum_{i=1}^3 \frac{\eta(\tau)^6 \theta_1(\tau, a_i) \theta_1(\tau, 2a_i-4\epsilon_+)}{\theta_1(\tau, \epsilon_1) \theta_1(\tau, \epsilon_2) \prod_{j\neq i} \theta_1(\tau, a_i-a_j) \theta_1(\tau, 2\epsilon_++a_j) \theta_1(\tau, 2\epsilon_+-a_i+a_j)} \, ,
\end{align}
where $ \theta_1(\tau,x) $ is the Jacobi theta function defined in Appendix~\ref{app:special}, and $ a_j = \langle \rho_j, \phi \rangle $ for $ \rho_j \in w(\mathbf{3}) $, i.e., $ (a_1,a_2,a_3) = (\phi_1,-\phi_1+\phi_2,-\phi_2) $. For $ k>1 $, the linear system is invertible, and therefore the blowup equations determine $ Z_k $ recursively from the lower-string elliptic genera.

\subsection{Codimension-4 defects and Wilson surfaces}

We now introduce codimension-4 defects realized by 2d chiral fermions coupled to the bulk $ \su 3 $ gauge symmetry in the fundamental representation $ \mathbf{3} $ or in the anti-fundamental representation $ \overline{\mathbf{3}} $. The 1-loop contributions \eqref{eq:X0} of these 2d fermions are
\begin{align}\label{eq:su3-x0}
    \mathcal{X}_0^{(+)} = \prod_{j=1}^3 \frac{\theta_1(\tau, x + a_j)}{\eta(\tau)} \, , \quad
    \mathcal{X}_0^{(-)} = \prod_{j=1}^3 \frac{\theta_1(\tau, x - a_j)}{\eta(\tau)} \, ,
\end{align}
where $ (a_1,a_2,a_3)=(\phi_1,-\phi_1+\phi_2,-\phi_2) $ as above and $ x $ is the holonomy of the $ \U{1}_x $ flavor symmetry acting on the 2d fermions. Since $ b=1 $ and $ A_{\mathbf{3}}=A_{\bar{\mathbf{3}}} = 1/2 $ in the anomaly cancellation condition \eqref{eq:anomaly-codim4}, we obtain $ \mathcal{E}_1 = \varphi $. The full defect partition functions therefore take the form
\begin{align}
    \mathcal{X}^{(\pm)} = e^{-2\pi i \mathcal{E}_0} \cdot Z_{\mathrm{pert}} \cdot \sum_{k=0}^\infty Q^{3k-1} \mathcal{X}_k^{(\pm)} \, .
\end{align}
where $ Q^{-1} = e^{-2\pi i \mathcal{E}_1} $ is the contribution of the defect gauge anomaly cancellation. Here, $ \mathcal{X}_k^{(\pm)} $ is the $ k $-string elliptic genus in the presence of the codimension-4 defect, and $ \mathcal{X}_0^{(\pm)} $ is the 1-loop contribution \eqref{eq:su3-x0}. The defect partition functions satisfy the blowup equations with the set of magnetic fluxes
\begin{align}\label{eq:su3-codim4-flux}
    \ell \in \mathbb{Z} + \nu \, , \ (\nu = \pm 1/6) \, , \quad
    n_i \in \mathbb{Z} \, , \quad
    B_\tau = B_x = 0 \, ,
\end{align}
where no background magnetic flux is turned on for the defect parameter $ x $. After summing over $ \ell $, the blowup equation at order $ Q^{3k-1} $ reads
\begin{align}\label{eq:su3-codim4-blowup}
    \thtftn{-\nu}{1/2}(3\tau, 2\epsilon_+ + \epsilon_1) \mathcal{X}_k^{(\pm)} = \sum_{P_k} (-1)^{n_1+n_2} \thtftn{-\nu}{1/2}(3\tau, 2\epsilon_++\epsilon_1-\mu) \mathcal{A}(n) \mathcal{X}_{k_1}^{(\pm,N)} Z_{k_2}^{(S)} \, ,
\end{align}
where $ \mu $, $ P_k $ and $ \mathcal{A}(n) $ are given in \eqref{eq:su3-mu}, \eqref{eq:su3-P} and \eqref{eq:su3-A0}, and $ \mathcal{X}_k^{(\pm, N)} $ is defined from $ \mathcal{X}_k^{(\pm)} $ in the same way as $ Z_k^{(N)} $ from $ Z_k $, with $ x $ unshifted. At this order, the two blowup equations with $ \nu=\pm 1/6 $ form a linear system for the two unknowns $ \mathcal{X}_k^{(\pm)} $ and $ \mathcal{X}_k^{(\pm,N)} $. This linear system is non-degenerate at $ k=1 $, and for $ k>1 $. Hence, the blowup equations determine $ \mathcal{X}_k^{(\pm)} $ recursively starting from the perturbative contribution \eqref{eq:su3-x0} and the bulk elliptic genera $ Z_k $.

As discussed in section~\ref{sec:codim4}, these codimension-4 defect partition functions can be decomposed into the expectation values of the three fundamental Wilson surfaces. The fundamental Wilson surfaces are labeled by the finite parts $ \lambda_0=0 $, $ \lambda_1=\omega_1 $ and $ \lambda_2=\omega_2 $ of the integrable highest weights of $ \widehat{\mathfrak{su}}(3)_1 $, where $ \omega_i $ are the fundamental weights of the $ \mathfrak{su}(3) $ algebra. At the perturbative level, the Wilson surface expectation values are the corresponding level-one affine characters,
\begin{align}
    W_j^{(0)} = \frac{1}{\eta(\tau)^2} \sum_{n \in \Lambda_r} q^{\frac{1}{2}\langle n+\lambda_j, n+\lambda_j \rangle} e^{2\pi i \langle n+\lambda_j, \phi \rangle} \, ,
\end{align}
where $ \Lambda_r $ is the root lattice. Explicitly, they can be written as
\begin{align}
    \begin{aligned}
        W_0^{(0)} &= \frac{1}{\eta(\tau)^2} \sum_{n_i \in \mathbb{Z}} q^{\frac{1}{2}\langle n, n \rangle} e^{2\pi i \langle n, \phi \rangle} = q^{-1/12} \left( 1 + \mathcal{O}(q) \right) \, , \\
        W_1^{(0)} &= \frac{q^{1/3}}{\eta(\tau)^2} \sum_{n_i \in \mathbb{Z}} q^{\frac{1}{2}\langle n, n \rangle + n_1} e^{2\pi i \langle n, \phi \rangle} e^{2\pi i \phi_1} = q^{1/4} \biggr( \sum_{j=1}^3 e^{2\pi i a_j} + \mathcal{O}(q) \biggl) \, , \\
        W_2^{(0)} &= \frac{q^{1/3}}{\eta(\tau)^2} \sum_{n_i \in \mathbb{Z}} q^{\frac{1}{2}\langle n, n \rangle + n_2} e^{2\pi i \langle n, \phi \rangle} e^{2\pi i \phi_2} = q^{1/4} \biggl( \sum_{j=1}^3 e^{-2\pi i a_j} + \mathcal{O}(q) \biggr) \, ,
    \end{aligned}
\end{align}
whose leading terms are the characters of the $ \mathbf{1} $, $ \mathbf{3} $ and $ \overline{\mathbf{3}} $ of $ \su 3 $. Unlike the codimension-4 defect partition function \eqref{eq:su3-x0}, these Wilson surface VEVs do not depend on the defect parameter $ x $. The perturbative defect partition functions \eqref{eq:su3-x0} decompose as
\begin{align}\label{eq:su3-codim4-decomp}
    \mathcal{X}_0^{(\pm)}(\phi, \tau, x) &= \sum_{j=0}^2 C_{\pm j}(\tau, x) W_j^{(0)}(\phi, \tau) \, ,
\end{align}
where the index $ j $ is understood modulo 3, and the coefficients $ C_j(\tau, x) $ are defined as 
\begin{gather}
    \begin{gathered}
        C_0(\tau, x) = \frac{1}{\eta(\tau)} \thtftn{1/2}{1/2}(3\tau, 3x) \, , \quad
        C_1(\tau, x) = \frac{e^{2\pi i/3}}{\eta(\tau)}\thtftn{-1/6}{1/2}(3\tau, 3x) \, , \\
        C_2(\tau, x) = \frac{e^{4\pi i/3}}{\eta(\tau)} \thtftn{1/6}{1/2}(3\tau, 3x) \, .
    \end{gathered}
\end{gather}
They are independent of the gauge holonomy $ \phi $.

The VEV of the fundamental Wilson surfaces including the instanton string corrections are given by
\begin{align}
    W_j = \frac{\mathcal{W}_j}{Z} = \sum_{k=0}^\infty Q^{3k-1} W_j^{(k)} \, ,
\end{align}
where we include the $ Q^{-1} $ factor that records the affine level as in \eqref{eq:wilsonVEV}. The full partition function $ \mathcal{W}_j $, which is expanded as
\begin{align}\label{eq:su3-W-expand}
    \mathcal{W}_j = e^{-2\pi i \mathcal{E}_0} \cdot Z_{\mathrm{pert}} \cdot \sum_{k=0}^\infty Q^{3k-d_i^\vee} \mathcal{W}_j^{(k)} \, ,
\end{align}
is also a solution of the blowup equations \eqref{eq:su3-codim4-blowup}, with the fluxes \eqref{eq:su3-codim4-flux} and $ \mathcal{X} $ replaced by $ \mathcal{W}_j $. The decomposition \eqref{eq:su3-codim4-decomp} extends to all orders in $ Q $, since the blowup equations are linear and no background fluxes are turned on for $ \tau $ and $ x $ in the blowup equation \eqref{eq:su3-codim4-blowup}. The coefficients $ C_j(\tau,x) $ are not shifted and factor out of the recursion. Thus,
\begin{align}
    \frac{\mathcal{X}^{(\pm)}}{Z} = \sum_{j=0}^2 C_{\pm j}(\tau, x) W_j \, ,
\end{align}
where $ Z $ is the partition function \eqref{eq:su3-fullZ} without a defect, and $ W_j $ does not depend on the defect parameter $ x $.

\subsection{Codimension-2 defects}

We next consider a codimension-2 defect supported on $ \mathbb{C}_{\epsilon_1} \times T^2 $, realized by coupling 4d $ \mathcal{N}=1 $ chiral multiplets to the bulk $ \su 3 $ gauge symmetry. By the cubic gauge anomaly cancellation condition discussed in section~\ref{sec:codim2}, a single chiral multiplet in the fundamental representation $ \mathbf{3} $ of the bulk $ \su 3 $ gauge symmetry is not allowed. We therefore consider the defect in the $ \mathbf{3} \oplus \bar{\mathbf{3}} $ representation, with both chiral multiplets having charge $ +1 $ under the $ \U{1}_x $ defect flavor symmetry. The anomaly cancellation condition \eqref{eq:codim2-anomaly} requires
\begin{align}
    \mathcal{E}_2^{\text{tree}} = \frac{2x\varphi}{\epsilon_1} \, .
\end{align}
Using \eqref{eq:psi-pert2}, the 1-loop contribution can be written as
\begin{align}
    \Psi_{\mathrm{pert}} &= \PE\left[ \frac{\sqrt{p_1}}{(1-p_1)(1-q)} \sum_{j=1}^3 \left( e^{2\pi i x} - e^{-2\pi i x} \right) \left( e^{2\pi i a_j} + q e^{-2\pi i a_j} \right) \right] \, ,
\end{align}
and there is no 1-loop contribution to the defect prepotential: $ \mathcal{E}_2 = \mathcal{E}_2^{\text{tree}} $. If one instead uses the 1-loop partition function \eqref{eq:psi-pert1}, there is a non-trivial contribution to the effective prepotential given by $ \mathcal{E}_2^{\text{1-loop}} = \langle \phi, \phi \rangle/(2\epsilon_1) $. However, this choice does not affect the instanton string correction of the defect partition function.

The full codimension-2 defect partition function is given by
\begin{align}\label{eq:su3-Psi}
    \Psi = e^{-2\pi i (\mathcal{E}_0 + \mathcal{E}_2)} \cdot Z_{\mathrm{pert}} \cdot \Psi_{\mathrm{pert}} \cdot \left( 1 + \sum_{k=1}^\infty Q^{3k} \Psi_k \right) \, .
\end{align}
This partition function satisfies the blowup equations constructed from the set of magnetic fluxes
\begin{align}
    \ell \in \mathbb{Z} + \nu \, , \ (\nu = \pm 1/6) \, , \quad
    n_i \in \mathbb{Z} \, , \quad
    B_\tau = B_x = 0 \, .
\end{align}
In particular, the $ k $-th order of the blowup equations reads
\begin{equation}
    \thtftn{-\nu}{1/2}(3\tau, 2\epsilon_+ + 2x) \Psi_k = \sum_{P_k} (-1)^{n_1+n_2} \thtftn{-\nu}{1/2}(3\tau, 2\epsilon_++2x-\mu) \mathcal{A}(n) \mathcal{D}(n) \Psi_{k_1}^{(N)} Z_{k_2}^{(S)} \, ,
\end{equation}
where $ \mu $ and $ P_k $ are given in \eqref{eq:su3-mu} and \eqref{eq:su3-P}. The 1-loop contribution of the defect $ \mathcal{D}(n) $ is given by
\begin{align}
    \mathcal{D}(n) = \prod_{\rho\in w(\mathbf{3})} \prod_{j=0}^{|\langle \rho, n \rangle|-1} \frac{\theta_1\bigl(\tau, s_\rho x + \langle \rho, \phi \rangle + s_\rho(j+\frac{1}{2}) \epsilon_1\bigr)}{\theta_1\bigl(\tau, -s_\rho x + \langle \rho, \phi \rangle + s_\rho(j+\frac{1}{2}) \epsilon_1\bigr)} \, .
\end{align}
where $ s_\rho = \sgn \langle \rho, n \rangle $. At order $ k $, there are two unknowns $ \Psi_k $ and $ \Psi_k^{(N)} $ in the blowup equations. The two equations with $ \nu=\pm 1/6 $ form a linear system for them, and it can be inverted to uniquely determine the elliptic genus $ \Psi_k $ recursively in $ k $ from the bulk elliptic genera $ Z_k $. One can check that using the other choice of the 1-loop partition function \eqref{eq:psi-pert1} and corresponding effective prepotential gives the same elliptic genera $ \Psi_k $ by explicitly solving the blowup equation.

\subsection{Seiberg--Witten curve} \label{sec:su3-sw}

In this subsection, we propose the quantum SW curve of the 6d $ \su 3 $ gauge theory using the defect partition functions. Let us define the wavefunction $ \psi(x) $ as the NS-limit of the normalized codimension-2 defect partition function with the $ x\to x-\epsilon_1/2 $ shift:
\begin{align}
    \psi(x; \varphi, \phi, \tau, \epsilon) = \lim_{\epsilon_2 \to 0} \frac{\Psi(\varphi, \phi, \tau, x-\frac{\epsilon_1}{2}, \epsilon_1, \epsilon_2)}{Z(\varphi, \phi, \tau, \epsilon_1, \epsilon_2)} \bigg|_{\epsilon_1=\epsilon} \, .
\end{align}
Its perturbative part is
\begin{align}
    \psi_{\mathrm{pert}}(x) = \prod_{j=1}^3 \Gamma_e(X v_j; p,q) \Gamma_e(q X/v_j; p,q) = \prod_{j=1}^3 \frac{\Gamma_e(X v_j; p,q)}{\Gamma_e(p X^{-1} v_j; p,q)} \, ,
\end{align}
where $ X=e^{2\pi i x} $, $ v_j=e^{2\pi i a_j} $, $ p=e^{2\pi i \epsilon} $, and the second equality follows from the identity $ \Gamma_e(z;p,q) \Gamma_e(pq/z;p,q)=1 $. We denote the normalized NS-limits of the codimension-4 defect partition functions by
\begin{align}
    \chi_\pm(x; \varphi, \phi, \tau, \epsilon) &= \lim_{\epsilon_2 \to 0} \frac{\mathcal{X}^{(\pm)}(\varphi, \phi, \tau, x, \epsilon_1, \epsilon_2)}{Z(\varphi, \phi, \tau, \epsilon_1, \epsilon_2)} \bigg|_{\epsilon_1=\epsilon} \, .
\end{align}
Applying the shift operator $ Y $, defined by $ Yf(x) = f(x+\epsilon) $, to the perturbative wavefunction $ \psi_{\mathrm{pert}} $ gives
\begin{align}\label{eq:su3-curve-pert}
    \frac{Y e^{-2\pi i \mathcal{E}_2} \psi_{\mathrm{pert}}(x)}{e^{-2\pi i \mathcal{E}_2} \psi_{\mathrm{pert}}(x)} = Q^{-2} q^{-1/2} \left( \prod_{j=1}^3 \frac{\theta_1(\tau,x+a_j) \theta_1(\tau,x-a_j)}{\eta(\tau)^2} \right) \, ,
\end{align}
where the RHS corresponds, up to the factor $ q^{-1/2} $, to the leading order of the product of codimension-4 defect partition functions $ \chi_+ \chi_- $. The additional factor $ q^{-1/2} $ can be cancelled by introducing a background factor in the effective prepotential,
\begin{align}
    \mathcal{E}_2 \to \mathcal{E}_2 - \frac{\tau x}{2\epsilon_1} \, ,
\end{align}
which is independent of $ \phi $ and therefore does not affect the blowup equations and hence the elliptic genera $ \Psi_k $. We propose that, including the higher-string corrections, the wavefunction $ \psi(x) $ satisfies the difference equation
\begin{align}\label{eq:su3-curve}
    \left( Y - \left( A_+(x) \chi_+(x) + A_-(x) \chi_-(x-\epsilon) \right) Y^{-1} - B(x) Y^{-2} \right) \psi(x) = \chi_+(x) \chi_-(x) \psi(x) \, ,
\end{align}
where
\begin{align}\label{eq:su3-curveAB}
    A_+(x) &= \frac{\theta_1(\tau,x) \theta_1(\tau,2x-\epsilon) \theta_1(\tau,2x-2\epsilon)}{\eta(\tau)^3} \, , \quad
    A_-(x) = \frac{\theta_1(\tau,x-\epsilon) \theta_1(\tau,2x) \theta_1(\tau,2x-\epsilon)}{\eta(\tau)^3} \, , \nonumber \\
    B(x) &= \frac{\theta_1(\tau,x-\epsilon)^2 \theta_1(\tau,2x) \theta_1(\tau,2x-\epsilon) \theta_1(\tau,2x-3\epsilon) \theta_1(\tau,2x-4\epsilon)}{\eta(\tau)^6} \, ,
\end{align}
are independent of the dynamical parameters $ \varphi $ and $ \phi $ of the theory. Note that the RHS \eqref{eq:su3-curve-pert} can be interpreted as the perturbative partition function of the codimension-4 defect realized by 2d fermions in the $ \mathbf{3}+\bar{\mathbf{3}} $ representation, which has $ \mathcal{E}_1=2\varphi $. Its elliptic genera can be computed from the blowup equations. At the generic $ \Omega $-deformation parameters, its partition function differs from the product of those for the individual defects in $ \mathbf{3} $ and $ \bar{\mathbf{3}} $. In the NS-limit, however, it factorizes as $ \chi_+ \chi_- $.

The difference equation can also be written in the factorized form
\begin{align}\label{eq:su3-curve2}
    \Bigl[ Y - \bigl( \chi_+(x) + A_-(x) Y^{-1} \bigr) \bigl( \chi_-(x) + A_+(x) Y^{-1} \bigr) \Bigr] \psi(x) = 0 \, .
\end{align}
We have checked this difference equation up to 3-string order. It can also be expressed in terms of the Wilson surface VEVs. We denote the NS-limits of the Wilson surface VEVs by
\begin{align}
    U_{j}(\varphi, \phi, \tau, \epsilon) = \lim_{\epsilon_2 \to 0} W_{j}(\varphi, \phi,\tau,\epsilon_1,\epsilon_2) \Big|_{\epsilon_1=\epsilon} \, .
\end{align}
Then, they satisfy
\begin{align}
    \chi_\pm(x) = \sum_{j=0}^2 C_{\pm j}(\tau,x) U_j \, ,
\end{align}
where $ U_j $ does not depend on $ x $, while $ C_j $ does not depend on the dynamical parameters $ \varphi $ and $ \phi $ of the theory. Substituting this into \eqref{eq:su3-curve2}, the dynamical parameters enter the difference equation only through the three Wilson surface VEVs $ U_j $.

We propose \eqref{eq:su3-curve2} as the quantum SW curve of the 6d minimal $ \su 3 $ gauge theory. As a further consistency check, we compute its classical periods and compare the resulting genus-zero free energy with the elliptic genus. We define
\begin{align}
   y(x) = \lim_{\epsilon \to 0} \psi(x+\epsilon)/\psi(x) \, , \quad
   \chi_\pm^{\mathrm{cl}}(x) = \lim_{\epsilon\to 0} \chi_\pm(x)  \, .
\end{align}
Motivated by the perturbative codimension-4 defects, we parametrize
\begin{align}
    \chi_\pm^{\mathrm{cl}} = \mathcal{Q}^{-1} P_\pm(x;z) \, , \quad
    P_\pm(x;z) = \prod_{j=1}^3 \frac{\theta_1(\tau, x\pm z_j)}{\eta(\tau)} \, , \quad
    \mathcal{Q} = e^{2\pi i \beta} \, ,
\end{align}
where $ \sum z_j = 0 $. The curve parameters $ z_j $ and $ \beta $ are related to the physical parameters $ a_j $ and $ \varphi $ by the periods below. The classical limit of the SW curve \eqref{eq:su3-curve2} is the cubic
\begin{align}\label{eq:su3-classical-sw}
    y^3 - \frac{P_+ P_-}{\mathcal{Q}^2} y^2 - \frac{A(x) ( P_+ + P_- )}{\mathcal{Q}} y - A(x)^2 = 0 \, , \quad
    A(x) = \frac{\theta_1(\tau,x) \theta_1(\tau,2x)^2}{\eta(\tau)^3} \, .
\end{align}
We define the logarithmic differential
\begin{align}\label{eq:SW-diff}
    \lambda = \log\left( \frac{\eta(\tau)^6 y(x)}{\theta_1(\tau,x)^6} \right) dx \, ,
\end{align}
which we use as the SW differential. We choose the branch $ y \sim \mathcal{Q}^{-2} P_+P_- $ as $ \mathcal{Q}\to 0 $, whose logarithm has the expansion
\begin{align}\label{eq:SW-logy}
    \log y = -4\pi i \beta + \log(P_+ P_-) + \mathcal{Q}^3 R(x;z) + \mathcal{O}(\mathcal{Q}^6) \, .
\end{align}
where
\begin{align}
    R(x;z) = \frac{A(x) (P_+ + P_-)}{P_+^2 P_-^2}
    \overset{x\to z_j}{=} \frac{\rho_j(z)}{(x-z_j)^2} + \frac{c_j(z)}{x-z_j} + \mathcal{O}(1) \, .
\end{align}
The last expression is the expansion $ R(x;z) $ near $ x=z_j $, where the double-pole coefficient $ \rho_j $ is given by
\begin{align}
    \rho_j(z) = \frac{A(z_j)}{[P_-'(z_j; z)]^2 P_+(z_j; z)} \, .
\end{align}
Here, the prime on $ P_- $ denotes differentiation with respect to its first argument. Since $ R $ is an even function of $ x $, its poles at $ x=-z_j $ have double-pole coefficients $ \rho_j $ and simple-pole coefficient $ -c_j $. With $ f(x) \equiv \partial_x \log \theta_1(\tau,x) $, matching the poles at $ x=\pm z_j $ gives
\begin{equation}
    R(x;z) = -\sum_{j=1}^3 \Bigl[ \rho_j(z) ( f'(x-z_j) + f'(x+z_j) ) - c_j(z) ( f(x-z_j) - f(x+z_j) ) \Bigr] + R_0(z) .
\end{equation}
Since $ f(x) \sim 1/x $ and $ f'(x) \sim -1/x^2 $ near $ x=0 $, this expression reproduces all the poles of $ R $. Subtracting these terms leaves an elliptic function, i.e., $ R_0(x+1;z)=R_0(x+\tau;z)=R_0(x;z) $, without poles, which is therefore independent of $ x $. We denote this remainder by $ R_0(z) $.

We now define the period integrals of the SW differential \eqref{eq:SW-diff} as
\begin{align}
    a_j = \frac{1}{2\pi i} \oint_{\mathcal{A}_j} x d\log y \, , \quad
    \Pi_A = \oint_{A} \lambda \, , \quad
    \Pi_B = \oint_{B} \lambda \, ,
\end{align}
where $ \mathcal{A}_j $ is a small loop around $ x=z_j $, and $ A $ and $ B $ run from a base point $ x_0 $ to $ x_0+1 $ and $ x_0+\tau $, respectively. We choose the logarithm branches by continuation from $ z_j=0 $ and deform the paths without crossing zeros or poles. At zero string order, these periods are
\begin{align}
    a_j = z_j + \mathcal{O}(\mathcal{Q}^3) , \quad
    \Pi_A = -4\pi i \beta + \mathcal{O}(\mathcal{Q}^3) , \quad
    \Pi_B = -4\pi i \tau \beta - 2\pi i \sum_j z_j^2 + \mathcal{O}(\mathcal{Q}^3) ,
\end{align}
where we use the quasi-periodicity of the theta function \eqref{eq:tht1-periodic}. These leading-order expressions suggest the identification
\begin{align}\label{eq:period-identify}
    \varphi = -\frac{\Pi_A}{4\pi i} = \beta + \mathcal{O}(\mathcal{Q}^3) \, , \quad
    \partial_\varphi \mathcal{F}_0 = \frac{3}{2} \Pi_B = -6\pi i \tau \varphi - 3\pi i \sum_j a_j^2 + \mathcal{O}(\mathcal{Q}^3) \, ,
\end{align}
where $ \mathcal{F}_0 $ is the genus-zero free energy
\begin{align}
    \mathcal{F}_0 = \lim_{\epsilon_1,\epsilon_2 \to 0} \epsilon_1 \epsilon_2 \log Z \, .
\end{align}
The normalization in \eqref{eq:period-identify} is fixed by the tree-level prepotential \eqref{eq:su3-tree}.

At one-string order, \eqref{eq:SW-logy} gives the corrections to the gauge and tensor parameters as
\begin{align}
    \begin{aligned}\label{eq:period-A}
        a_j &= z_j + \frac{\mathcal{Q}^3}{2\pi i} \oint_{\mathcal{A}_j} x \, dR + \mathcal{O}(\mathcal{Q}^6) = z_j - \mathcal{Q}^3 c_j(z) + \mathcal{O}(\mathcal{Q}^6) \, , \\
        \varphi &= \beta - \frac{\mathcal{Q}^3}{4\pi i} \oint_A R\, dx + \mathcal{O}(\mathcal{Q}^6) = \beta - \frac{\mathcal{Q}^3 R_0(z)}{4\pi i} + \mathcal{O}(\mathcal{Q}^6) \, ,
    \end{aligned}
\end{align}
where the second line follows from $ f(x+1)=f(x) $ and $ f(x+\tau)=f(x)-2\pi i $. Similarly, these relations give
\begin{align}
    \Pi_B &= -4\pi i \tau \beta - 2\pi i \sum_j z_j^2 + \mathcal{Q}^3 \biggl( 4\pi i \sum_j (\rho_j(z) + z_j c_j(z) ) + \tau R_0(z) \biggr) + \mathcal{O}(\mathcal{Q}^6) \, .
\end{align}
Inverting \eqref{eq:period-A} and substituting into $ \Pi_B $ gives
\begin{align}
    \partial_\varphi \mathcal{F}_0 &= -6\pi i \tau \varphi - 3\pi i \sum_j a_j^2 + 6\pi i Q^3 \sum_j \rho_j(a) + \mathcal{O}(Q^6) \\
    &= -6\pi i \tau \varphi - 3\pi i \sum_j a_j^2 + 6\pi i Q^3 \sum_j \frac{\eta(\tau)^6 \theta_1(\tau,a_j) \theta_1(\tau,2a_j)}{\theta_1'(\tau,0)^2 \prod_{k\neq j} \theta_1(\tau,a_j-a_k)^2 \theta_1(\tau,a_k)}  + \mathcal{O}(Q^6) \, . \nonumber
\end{align}
The $ Q^3 $ coefficient exactly matches  the genus-zero limit of the 1-string elliptic genus \eqref{eq:su3-Z1}. This provides a non-trivial one-string check for the proposed SW geometry \eqref{eq:su3-curve2}.

\section{Higgsable 6d SCFTs on a \texorpdfstring{$-3$}{-3} curve} \label{sec:higgsable}

In this section, we propose quantum SW curves for 6d $ \so N $ and $ \mathfrak{g}_2 $ gauge theories on a $ -3 $ curve, as well as for the $ \su 3 $ theory compactified on a circle with a $ \mathbb{Z}_2 $ outer-automorphism twist. The $ \so N $ and $ \mathfrak{g}_2 $ theories lie on a Higgsing tree whose endpoint is pure $ \su 3 $ gauge theory. We consider further consistency checks of the SW curve of the pure $ \su 3 $ gauge theory by taking the corresponding Higgs branch limits of the proposed curves. We also study another branch flow from the $ \mathfrak{g}_2 $ theory that yields the twisted compactification of the $ \su 3 $ theory.

\subsection{\texorpdfstring{$ \so N $}{so(N)} gauge theory} \label{sec:so}

In this subsection, we consider defects and quantum SW curves of 6d $ \so N $ gauge theories on a $ -3 $ curve. We write $ r=\rk(\so N) = \lfloor N/2 \rfloor $ in this subsection. The matter content of each theory is given by
\begin{alignat}{3}
    \begin{aligned}
        &\so7 \ : \ 2 \times \mathbf{8} \, , \quad
        &&\so8 \ : \ \mathbf{8}_v + \mathbf{8}_s + \mathbf{8}_c \, , \quad
        &&\so9 \ : \ 2 \times \mathbf{9} + \mathbf{16} \, , \\
        &\so{10} \ : \ 3 \times \mathbf{10} + \mathbf{16} \, ,
        &&\so{11} \ : \ 4 \times \mathbf{11} + \frac{1}{2}\mathbf{32} \, ,
        &&\so{12} \ : \ 5 \times \mathbf{12} + \frac{1}{2}\mathbf{32} \, ,
    \end{aligned}
\end{alignat}
where $ \frac{1}{2} $ denotes a half-hypermultiplet, $ \so 8 $ has three 8-dimensional representations, which are the vector $ \mathbf{8}_v $, spinor $ \mathbf{8}_s $ and conjugate spinor $ \mathbf{8}_c $. The case $ N>12 $ is not allowed for a theory supported on a single $ -3 $ curve \cite{Heckman:2015bfa}. It is possible to construct the blowup equations for these theories. However, an unpaired half-hypermultiplet forces $ \Lambda=0 $ in the blowup equations \eqref{eq:blowup} for 6d theories \cite{Gu:2020fem, Kim:2020hhh}, which makes it difficult to compute the elliptic genera from the blowup equations. For this reason, we restrict our analysis to $ N \leq 10 $.

\paragraph{Without a defect}

The tree-level effective prepotential of the $ \so N $ theory is given by
\begin{align}
    \epsilon_1 \epsilon_2 \mathcal{E}_0^{\mathrm{tree}}
    &= \frac{3}{2}\tau \varphi^2 + 3 \varphi \left( -\frac{\epsilon_1^2 + \epsilon_2^2}{12} + \frac{1}{2} \langle \phi, \phi \rangle - \sum_{f} \frac{A_{\mathbf{R}_f}}{3} m_f^2 + \frac{N-2}{3} \epsilon_+^2 \right) \, ,
\end{align}
where $ \varphi $ is the tensor parameter, $ \langle \phi, \phi \rangle $ is the invariant quadratic form of the $ \so N $ gauge holonomy $ \phi=\sum_{i=1}^{r} \phi_i \alpha_i^\vee $ with simple coroots $ \alpha_i^\vee $ of the gauge algebra, and $ A_{\mathbf{R}_f} $ is the quadratic index of the representation $ \mathbf{R}_f $ of the $ f $-th hypermultiplet. The full partition function on $ \mathbb{R}^4 \times T^2 $ without a defect takes the form \eqref{eq:su3-fullZ}. This partition function satisfies the blowup equations constructed from the set of magnetic fluxes
\begin{align}\label{eq:so-flux}
    \ell \in \mathbb{Z} + \nu \, , \ (\nu = \pm 1/6, 1/2) \, , \quad
    n_i \in \mathbb{Z} \, , \quad
    B_\tau = 0 \, , \quad
    B_{m_f} = 1/2 \, ,
\end{align}
where $ \ell $ and $ n = \sum n_i \alpha_i^\vee $ are dynamical magnetic fluxes for the tensor parameter $ \varphi $ and gauge parameters $ \phi $, respectively, and $ B_{m_f} $ is the background magnetic flux associated with $ m_f $. For each set of magnetic fluxes, the blowup equation at $ k $-string order is given by
\begin{align}\label{eq:so-blowup}
    \thtftn{-\nu}{1/2}(3\tau, \lambda) Z_k = \sum_{P_k} \varepsilon(n) \thtftn{-\nu}{1/2}(3\tau, \lambda - \mu) \mathcal{A}(n) Z_{k_1}^{(N)} Z_{k_2}^{(S)} \, ,
\end{align}
where $ \varepsilon(n) = (-1)^{|n| + \frac{1}{2}(N-7) \langle n, n \rangle} $ with $ |n| = \sum n_i $, and $ P_k $ is the set of tuples $ (n_i, k_1, k_2) $ defined as \eqref{eq:su3-P}, with the symmetric bilinear form $ \langle \cdot, \cdot \rangle $ of $ \so N $. We define $ \lambda $ and $ \mu $ by
\begin{align}
    \lambda = (N-3) \epsilon_+ + \sum_f A_{\mathbf{R}_f} m_f \, , \quad
    \mu = 3k_1 \epsilon_1 + 3k_2 \epsilon_2 + 3 \langle n, n \rangle \epsilon_+ + 3\langle n, \phi \rangle \, .
\end{align}
The 1-loop factor factorizes into vector and hypermultiplet contributions as
\begin{align}
    \mathcal{A}(n) = \mathcal{A}_{\mathrm{vec}}(n) \mathcal{A}_{\mathrm{hyp}}(n) \, ,
\end{align}
where $ \mathcal{A}_{\mathrm{vec}}(n) $ is given by \eqref{eq:su3-A0} for the set of positive roots $ \Delta^+ $ of $ \so N $, and
\begin{equation}\label{eq:so-A}
    \mathcal{A}_{\mathrm{hyp}}(n) = \prod_f \prod_{\rho \in w(\mathbf{R}_f)} \prod_{j_1=0}^{|n_{f\rho}| - \frac{3}{2}} \prod_{j_2=0}^{j_1} \frac{\theta_1(\tau,  \langle \rho, \phi \rangle + m_f + s_{f\rho}( (j_2+\frac{1}{2}) \epsilon_1 + (j_1-j_2+\frac{1}{2}) \epsilon_2 ))}{\eta(\tau)} ,
\end{equation}
where $ n_{f\rho} = \langle \rho, n \rangle + B_{m_f} $ and $ s_{f\rho} = \sgn n_{f \rho} $. The $ k $-string elliptic genus $ Z_k $ is determined by solving the three blowup equations with $ \nu=\pm 1/6, 1/2 $ for the three unknowns $ Z_k $, $ Z_k^{(N)} $ and $ Z_k^{(S)} $ at each $ k $. For the $ \so 7 $ gauge theory, the result agrees with the ADHM-like calculation in \cite{Kim:2018gjo}.

\paragraph{Codimension-4 defect}

Consider the codimension-4 defect realized by 2d fermions transforming in the vector representation $ \mathbf{N} $ of the bulk $ \so N $ gauge symmetry. The defect contribution to the effective prepotential and the perturbative defect partition function are
\begin{align}\label{eq:so-X0}
    \mathcal{E}_1 = 2\varphi \, , \quad
    \mathcal{X}_{0} = \prod_{j=1}^r \frac{\theta_1(\tau, x+a_j) \theta_1(\tau, x-a_j)}{\eta(\tau)^2} \, , 
\end{align}
where $ a_j $ are the $ \so N $ holonomies in an orthonormal basis, defined as
\begin{align}\label{eq:so-orthonormal}
    \so{2r} \ &: \ a_j = (\phi_1, -\phi_1+\phi_2, \cdots, -\phi_{r-3}+\phi_{r-2}, -\phi_{r-2} + \phi_{r-1} + \phi_r, -\phi_{r-1}+\phi_r)_j \, , \nonumber \\
    \so{2r+1} \ &: \ a_j = (\phi_1, -\phi_1+\phi_2, \cdots, -\phi_{r-2}+\phi_{r-1}, -\phi_{r-1}+2\phi_r)_j \, .
\end{align}
For odd $ N $, we omit the Coulomb-independent factor associated with the zero weight of the vector representation. The full defect partition function has the expansion
\begin{align}
    \mathcal{X} = e^{-2\pi i \mathcal{E}_0} \cdot Z_{\mathrm{pert}} \cdot \sum_{k=0}^\infty Q^{3k-2} \mathcal{X}_k \, .
\end{align}
The blowup equations for the full partition function $ \mathcal{X} $ can be constructed from the set of magnetic fluxes
\begin{align}\label{eq:so-codim4-flux}
    \ell \in \mathbb{Z} + \nu \, , \ (\nu = \pm 1/6) \, , \quad
    n_i \in \mathbb{Z} \, , \quad
    B_{m_f} = 1/2 \, , \quad
    B_\tau = B_x = 0 \, .
\end{align}
After summing over $ \ell $, the elliptic genus $ \mathcal{X}_k $ satisfies the blowup equations
\begin{equation}
    \thtftn{-\nu}{1/2}(3\tau, \lambda+2\epsilon_1) \mathcal{X}_k = \sum_{P_k} \varepsilon(n) \thtftn{-\nu}{1/2}(3\tau, \lambda-\mu+2\epsilon_1) \mathcal{A}(n) \mathcal{X}_{k_1}^{(N)} Z_{k_2}^{(S)} .
\end{equation}
Since there are two blowup equations distinguished by $ \nu=\pm 1/6 $, for the two unknowns $ \mathcal{X}_k $ and $ \mathcal{X}_k^{(N)} $ at each string order, these equations determine the defect elliptic genera $ \mathcal{X}_k $ recursively from the bulk elliptic genera.

\paragraph{Wilson surfaces}
Next, we consider the fundamental Wilson surface operators in $ \so N $ gauge theories. There are $ r+1 $ fundamental Wilson surfaces, labeled by the affine nodes $ j=0,1,\cdots,r $. At the perturbative level, their VEVs $ W_j^{(0)} $ are the affine characters with the highest weights $ \hat{\omega}_j $. Their levels are the dual Coxeter labels $ d_j^\vee $, as in section~\ref{sec:codim4}. When $ N=2r $, the four Wilson surfaces $ \mathcal{W}_j $ with $ j=0,1,r-1,r $ have level one, while the others have level two. For $ N=2r+1 $, the Wilson surfaces with $ j=0,1,r $ have level one, and the others have level two. Let us define
\begin{align}\label{eq:so-P}
    P_\mu(\tau, \phi) = \prod_{j=1}^r \frac{\theta_\mu(\tau, a_j)}{\eta(\tau)} \, .
\end{align}
Then, the level-one perturbative Wilson surface VEVs for $ \so{2r} $ are
\begin{alignat}{2}
    \begin{aligned}
        &W_0^{(0)} = \frac{1}{2} (P_3+P_4) \, , \ \quad
        &&W_1^{(0)} = \frac{1}{2} (P_3 - P_4) \, , \\
        &W_{r-1}^{(0)} = \frac{1}{2} (P_2 - i^{r} P_1) \, , \quad
        &&W_r^{(0)} = \frac{1}{2} (P_2 + i^{r} P_1) \, .
    \end{aligned}
\end{alignat}
For $ \so{2r+1} $, they are
\begin{gather}
    \begin{gathered}
        W_0^{(0)} = \frac{1}{2} \left( \sqrt{\frac{\theta_3(\tau,0)}{\eta(\tau)}} P_3 + \sqrt{\frac{\theta_4(\tau,0)}{\eta(\tau)}} P_4 \right) \, , \\
        W_1^{(0)} = \frac{1}{2} \left( \sqrt{\frac{\theta_3(\tau,0)}{\eta(\tau)}} P_3 - \sqrt{\frac{\theta_4(\tau,0)}{\eta(\tau)}} P_4 \right) \, , \quad
        W_r^{(0)} = \sqrt{\frac{\theta_2(\tau,0)}{2\eta(\tau)}} P_2 \, .
    \end{gathered}
\end{gather}
The level-two characters with $ 2\leq j \leq r-2 $ for even $ N $, or $ 2\leq j \leq r-1 $ for odd $ N $, are
\begin{align}
    W_j^{(0)} = \frac{1}{\eta(\tau)^{N-1}} \underset{|p|=j}{\sum_{p \in \mathbb{Z}^N}} q^{\frac{1}{2}\lVert p\rVert^2 - \frac{j^2}{2N}} e^{2\pi i p \cdot t} \, ,
\end{align}
where $ |p| = \sum_a p_a $, $ \lVert p \rVert^2 = p \cdot p = \sum_a p_a^2 $ and $ t=(a_1,-a_1,\cdots,a_r,-a_r) $ for even $ N $, with an additional zero entry for odd $ N $. The full partition functions $ \mathcal{W}_j $ in the presence of Wilson surface defects, with the string expansion \eqref{eq:su3-W-expand} at level $ d_j^\vee $, also satisfy the blowup equations. Using the magnetic fluxes \eqref{eq:so-flux}, we obtain the following blowup equations
\begin{equation}
    \thtftn{-\nu}{1/2}(3\tau, \lambda+d_j^\vee\epsilon_1) \mathcal{W}^{(k)}_j = \sum_{P_k} \varepsilon(n) \thtftn{-\nu}{1/2}(3\tau, \lambda-\mu+d_j^\vee\epsilon_1) \mathcal{A}(n) \mathcal{W}_{j}^{(k_1,N)} Z_{k_2}^{(S)} \, ,
\end{equation}
at order $ k $. The elliptic genus $ \mathcal{W}_j^{(k)} $ can be computed by solving the two equations with $ \nu=\pm 1/6 $.

\paragraph{Codimension-2 defect}

The partition function in the presence of a codimension-2 defect can be computed in a similar manner. We consider the defect realized by coupling a 4d chiral multiplet in the vector representation $ \mathbf{N} $ of the bulk $ \so N $ gauge symmetry, with charge $ +1 $ under $ \U{1}_x $. The defect contribution to the effective prepotential and the perturbative partition function are
\begin{align}
    \mathcal{E}_2 &= \frac{2x}{\epsilon_1} (\varphi + \phi_r ) \, , \quad
    \Psi_{\mathrm{pert}} = \PE\left[ \frac{\sqrt{p_1}\left( e^{2\pi i x} - e^{-2\pi i x} \right)}{(1-p_1)(1-q)} \sum_{j=1}^r  \left( e^{2\pi i a_j} + q e^{-2\pi i a_j} \right) \right] .
\end{align}
Here, we take $ \mathcal{S} $ in \eqref{eq:psi-pert2} to be the set of weights $ \rho $ satisfying $ \langle \rho, \phi \rangle = a_j $. The full codimension-2 defect partition function takes the form \eqref{eq:su3-Psi}. It satisfies the blowup equations constructed from the set of magnetic fluxes \eqref{eq:so-codim4-flux}. After summing over $ \ell $, the blowup equation at order $ k $ can be written as
\begin{align}
    \thtftn{-\nu}{1/2}(3\tau, \lambda+2x) \Psi_k = \sum_{P_k} \varepsilon(n) \thtftn{-\nu}{1/2}(3\tau, \lambda-\mu+2x) \mathcal{A}(n) \mathcal{D}(n) \Psi_{k_1}^{(N)} Z_{k_2}^{(S)} \, ,
\end{align}
where the 1-loop factor $ \mathcal{D}(n) $ is
\begin{align}\label{eq:so-D}
    \mathcal{D}(n) = \prod_{\rho\in \mathcal{S}} \prod_{j=0}^{|\langle \rho, n \rangle|-1} \frac{\theta_1\bigl(\tau, s_\rho x + \langle \rho, \phi \rangle + s_\rho(j+\frac{1}{2}) \epsilon_1\bigr)}{\theta_1\bigl(\tau, -s_\rho x + \langle \rho, \phi \rangle + s_\rho(j+\frac{1}{2}) \epsilon_1\bigr)} \, .
\end{align}
with $ s_\rho = \sgn \langle \rho, n \rangle $. Since there are two blowup equations with $ \nu=\pm 1/6 $ for the two unknown functions $ \Psi_k $ and $ \Psi_k^{(N)} $ at each order $ k $, the higher string elliptic genera are determined recursively from the bulk elliptic genera $ Z_k $.

\paragraph{Seiberg--Witten curve}

Using the explicit defect partition functions, we construct a quantum difference equation. As in section~\ref{sec:su3-sw}, we define the wavefunction $ \psi(x) $ as the NS-limit of the normalized codimension-2 defect partition function:
\begin{align}\label{eq:so-psi}
    \psi(x; \varphi, \phi, \tau, m,\epsilon) = \lim_{\epsilon_2 \to 0} \frac{\Psi(\varphi, \phi, \tau, x-\frac{\epsilon_1}{2}, m, \epsilon_1, \epsilon_2)}{Z(\varphi, \phi, \tau, m, \epsilon_1, \epsilon_2)} \bigg|_{\epsilon_1=\epsilon} \, ,
\end{align}
together with the shift $ \mathcal{E}_2 \to \mathcal{E}_2 - r\tau x/(6\epsilon_1) $, which is independent of the dynamical parameters of the theory. In this parametrization, the perturbative part of the wavefunction is given by
\begin{align}
    \psi_{\mathrm{pert}}(x) = \prod_{j=1}^r \frac{\Gamma_e(X v_j; p,q)}{\Gamma_e(pX^{-1}v_j; p,q)} \, ,
\end{align}
where $ X=e^{2\pi i x} $, $ v_j=e^{2\pi i a_j} $ and $ p=e^{2\pi i \epsilon} $. We also denote the NS-limits of the normalized codimension-4 defect partition functions by
\begin{align}\label{eq:so-codim4-ns}
    \begin{aligned}
    \chi(x; \varphi, \phi, \tau, m, \epsilon) &= \lim_{\epsilon_2 \to 0} \frac{\mathcal{X}(\varphi, \phi, \tau, x, m, \epsilon_1, \epsilon_2)}{Z(\varphi, \phi, \tau, m, \epsilon_1, \epsilon_2)} \bigg|_{\epsilon_1=\epsilon} \, , \\
    U_j(\varphi, \phi, \tau, m, \epsilon) &= \lim_{\epsilon_2 \to 0} W_j(\varphi, \phi, \tau, m, \epsilon_1, \epsilon_2) \Big|_{\epsilon_1=\epsilon} \, ,
    \end{aligned}
\end{align}
where we do not shift the parameter $ x $ for $ \mathcal{X} $. It is convenient to define
\begin{align}
    \mathcal{P}_1 = i^{-r}(U_r - U_{r-1}) \, , \quad
    \mathcal{P}_2 = U_{r-1} + U_r \, , \quad
    \mathcal{P}_3 = U_0 + U_1 \, , \quad
    \mathcal{P}_4 = U_0 - U_1 \, ,
\end{align}
for $ \so{2r} $ and
\begin{align}\label{eq:B-P}
    \mathcal{P}_2 = \sqrt{\frac{2\eta(\tau)}{\theta_2(\tau,0)}} U_r \, , \quad
    \mathcal{P}_3 = \sqrt{\frac{\eta(\tau)}{\theta_3(\tau,0)}} (U_0+U_1) \, , \quad
    \mathcal{P}_4 = \sqrt{\frac{\eta(\tau)}{\theta_4(\tau,0)}} (U_0-U_1) \, ,
\end{align}
for $ \so{2r+1} $. Their leading terms are $ \mathcal{P}_I = Q^{-1} P_I + \mathcal{O}(Q^2) $, with $ P_I $ defined in \eqref{eq:so-P}. We then find that the wavefunction $ \psi(x) $ satisfies
\begin{align}\label{eq:so-curve}
    \left( Y - \sum_{I} (-1)^I A_I(x) \mathcal{P}_I Y^{-1} - B(x) Y^{-2} \right) \psi(x) = \chi(x) \psi(x) \, ,
\end{align}
where
\begin{align}
    A_I(x) &= \frac{\theta_1(\tau, 2x-\epsilon)^2}{2\eta(\tau)^2} \Biggl[\prod_{i=1}^{N-7} \frac{\theta_1(\tau, x-\frac{\epsilon}{2} \pm m_i)}{\eta(\tau)^2}\Biggr] \Biggl[\prod_{J\neq I} \frac{\theta_J(\tau, x) \theta_J(\tau,x-\epsilon)}{\eta(\tau)^2} \Biggr] \Biggl[\prod_{z\in \mathcal{M}} \frac{\theta_I(\tau,z)}{\eta(\tau)} \Biggr] , \nonumber \\
    B(x) &= \frac{\theta_1(\tau,2x) \theta_1(\tau,2x-\epsilon) \theta_1(\tau,2x-3\epsilon) \theta_1(\tau,2x-4\epsilon)}{\eta(\tau)^4} \left[ \frac{\eta(\tau)^2}{\theta_1(\tau, x) \theta_1(\tau, x-2\epsilon)} \right]_{N=\mathrm{odd}} \nonumber \\
    &\quad \Biggl[\prod_{i=1}^{N-7} \frac{\theta_1(\tau,x-\frac{\epsilon}{2}\pm m_i) \theta_1(\tau,x-\frac{3\epsilon}{2}\pm m_i)}{\eta(\tau)^4} \Biggr] \Biggl[ \prod_{z\in \mathcal{M}^0} \frac{\theta_1(\tau, x-\epsilon \pm z)}{\eta(\tau)^2} \Biggr] \, , \label{eq:soN-curveB}
\end{align}
Here, the indices $ I $ and $ J $ run over $ 1,\cdots,4 $ for even $ N $ and $ 2,\cdots,4 $ for odd $ N $. The factor with the subscript $ N=\text{odd} $ in $ B(x) $ is included only for odd $ N $. We use the shorthand notation $ \theta_1(\tau,u\pm v) = \theta_1(\tau,u+v) \theta_1(\tau,u-v) $. The parameters $ m_i $, $ 1 \leq i \leq N-7 $, are the masses of the vector hypermultiplets of the $ \so N $ theory, and the set $ \mathcal{M} $ is defined in terms of the spinor masses by
\begin{alignat}{2}
    \begin{aligned}
        &\mathfrak{so}(7) \ : \ \{ 0, m_1\pm m_2\} \, , \quad
        &&\mathfrak{so}(8) \ : \ \{ m_2\pm m_3\} \, , \\
        &\mathfrak{so}(9) \ : \ \{ 0, 2m_3 \} \, , \quad
        &&\mathfrak{so}(10) \ : \ \{ -2m_4 \} \, .
    \end{aligned}
\end{alignat}
$ \mathcal{M}^0 $ is obtained by omitting the zero entry of $ \mathcal{M} $. We remark that the NS-limit of the normalized codimension-4 defect partition function $ \chi(x) $ can also be decomposed into quadratic products $ U_i U_j $ of level-one Wilson surface VEVs and linear terms in the level-2 Wilson surface VEVs, with coefficients depending only on $ \tau $ and $ x $. We have checked this decomposition up to 3-string order. After dividing by $ \psi(x) $, the three terms on the LHS of \eqref{eq:so-curve} begin at orders $ Q^{-2} $, $ Q^1 $, and $ Q^4 $ respectively, and we have checked this difference equation through order $ Q^7 $, corresponding to three-string order. These checks suggest that \eqref{eq:so-curve} is the quantum SW curve of $ \so N $ gauge theories on a $ -3 $ curve, for $ N \leq 10 $.

\subsection{\texorpdfstring{$ \mathfrak{g}_2 $}{g2} gauge theory} \label{sec:g2}

\paragraph{Without a defect}

In this subsection, we consider the $ \mathfrak{g}_2 $ gauge theory with one hypermultiplet in the fundamental representation $ \mathbf{7} $. The tree-level effective prepotential is given by
\begin{align}
    \epsilon_1 \epsilon_2 \mathcal{E}_0^{\mathrm{tree}}
    &= \frac{3}{2}\tau \varphi^2 + 3\varphi\left( -\frac{\epsilon_1^2 + \epsilon_2^2}{12} + \frac{1}{2} \langle \phi, \phi \rangle - \frac{m^2}{3} + \frac{4}{3} \epsilon_+^2 \right) \, ,
\end{align}
where $ \varphi $ is the tensor parameter, $ \phi = \phi_1 \alpha_1^\vee + \phi_2 \alpha_2^\vee $ is the $ \mathfrak{g}_2 $ gauge holonomy in the basis of simple coroots $ \alpha_i^\vee $, and $ m $ is the hypermultiplet mass parameter. The full partition function takes the form \eqref{eq:su3-fullZ} and satisfies the blowup equations constructed from the magnetic fluxes
\begin{align}
    \ell \in \mathbb{Z} + \nu \, , \ (\nu = \pm 1/6, 1/2) \, , \quad
    n_i \in \mathbb{Z} \, , \quad
    B_\tau = 0 \, , \quad
    B_m = 1/2 \, .
\end{align}
At order $ Q^{3k} $, the blowup equation reads
\begin{align}
    \thtftn{-\nu}{1/2}(3\tau, \lambda) Z_k = \sum_{P_k} (-1)^{n_1n_2} \thtftn{-\nu}{1/2}(3\tau, \lambda-\mu) \mathcal{A}(n) Z_{k_1}^{(N)} Z_{k_2}^{(S)} \, ,
\end{align}
where $ P_k $ is the set of tuples $ (n_i, k_1, k_2) $ defined in \eqref{eq:su3-P}, with the symmetric bilinear form $ \langle \cdot, \cdot \rangle $ of $ \mathfrak{g}_2 $. We define $ \lambda $ and $ \mu $ by
\begin{align}
    \lambda = 3\epsilon_+ + m \, , \quad
    \mu = 3k_1 \epsilon_1 + 3k_2 \epsilon_2 + 3\langle n, n \rangle \epsilon_+ + 3\langle n, \phi \rangle \, ,
\end{align}
and the 1-loop factor is $ \mathcal{A}(n) = \mathcal{A}_{\mathrm{vec}}(n) \mathcal{A}_{\mathrm{hyp}}(n) $, where $ \mathcal{A}_{\mathrm{vec}}(n) $ is given by \eqref{eq:su3-A0} for the positive roots of $ \mathfrak{g}_2 $ and $ \mathcal{A}_{\mathrm{hyp}}(n) $ is given by \eqref{eq:so-A} with $ \rho \in w(\mathbf{7}) $. The three blowup equations labeled by $ \nu $ form a linear system for $ Z_k $, $ Z_k^{(N)} $ and $ Z_k^{(S)} $ at each string order, and determine the $ k $-string elliptic genus $ Z_k $. The results agree with the elliptic genera computed from the ADHM-like description \cite{Kim:2018gjo} obtained by Higgsing $ \so 7 $ gauge theory. For instance, the 1-string elliptic genus is given by
\begin{align}
    Z_1 = \sum_{i=1}^3 \frac{\eta(\tau)^6 \theta_1(2a_i-4\epsilon_+) \theta_1(a_i - \epsilon_+ + m) \theta_1(a_i - \epsilon_+ - m)}{\theta_1(\epsilon_1) \theta_1(\epsilon_2) \theta_1(a_i-2\epsilon_+) \prod_{j\neq i} \theta_1(a_i-a_j) \theta_1(2\epsilon_++a_j) \theta_1(2\epsilon_+-a_i+a_j)} \, ,
\end{align}
where $ (a_1,a_2,a_3) = (\phi_1, \phi_1-\phi_2, -2\phi_1+\phi_2) $.

\paragraph{Codimension-4 defect}
We consider the codimension-4 defect realized by 2d chiral fermions transforming in the fundamental representation $ \mathbf{7} $ of the bulk $ \mathfrak{g}_2 $ gauge algebra. The defect contribution to the effective prepotential and the perturbative partition function are
\begin{align}
    \mathcal{E}_1 = 2\varphi \, , \quad
    \mathcal{X}_0 = \prod_{j=1}^3 \frac{\theta_1(\tau, x+a_j) \theta_1(\tau, x-a_j)}{\eta(\tau)^2} \, ,
\end{align}
where we omit the Cartan part of the fundamental representation. The full defect partition function has the expansion
\begin{align}
    \mathcal{X} &= e^{-2\pi i \mathcal{E}_0} \cdot Z_{\mathrm{pert}} \cdot \left( Q^{-2} \mathcal{X}_{0} + \sum_{k=1}^\infty Q^{3k-2} \mathcal{X}_k \right) \, .
\end{align}
The blowup equations for $ \mathcal{X} $ can be constructed from the set of magnetic fluxes
\begin{align}\label{eq:g2-codim4-flux}
    \ell \in \mathbb{Z} + \nu \, , \ (\nu = \pm 1/6) \, , \quad
    n_i \in \mathbb{Z} \, , \quad
    B_{\tau} = B_x = 0 \, , \quad
    B_{m} = 1/2 \, .
\end{align}
After summing over $ \ell $, the elliptic genus $ \mathcal{X}_k $ satisfies the blowup equations
\begin{align}
    \thtftn{-\nu}{1/2}(3\tau, \lambda+2\epsilon_1) \mathcal{X}_k = \sum_{P_k} (-1)^{n_1 n_2} \thtftn{-\nu}{1/2}(3\tau, \lambda-\mu+2\epsilon_1) \mathcal{A}(n) \mathcal{X}_{k_1}^{(N)} Z_{k_2}^{(S)} \, .
\end{align}
Since there are two blowup equations labeled by $ \nu=\pm 1/6 $ for the two unknowns $ \mathcal{X}_k $ and $ \mathcal{X}_k^{(N)} $ at each string order, we can determine the elliptic genera $ \mathcal{X}_k $ in the presence of the defect.

\paragraph{Wilson surfaces}

We next consider the three fundamental Wilson surface operators labeled by the affine nodes $ j=0,1,2 $ of $ \widehat{\mathfrak{g}}_2 $, given in Figure~\ref{fig:dynkin}. The perturbative VEVs $ W_j^{(0)} $ are the affine characters with integrable highest weights $ \hat{\omega}_j $. Their $ q $-expansions begin as
\begin{align}
    \begin{aligned}\label{eq:g2-affine}
        W_0^{(0)} &= q^{-7/60} \left( 1 + \chi_{\mathbf{14}} q + \left( 1 + \chi_{\mathbf{14}} + \chi_{\mathbf{27}} \right) q^2 + \mathcal{O}(q^3) \right) \, , \\
        W_1^{(0)} &= q^{17/60} \left( \chi_{\mathbf{7}} + (\chi_{\mathbf{7}} + \chi_{\mathbf{27}}) q + \mathcal{O}(q^2) \right) \, , \\
        W_2^{(0)} &= q^{17/36} \left( \chi_{\mathbf{14}} + \left( 1 + \chi_{\mathbf{14}} + \chi_{\mathbf{27}} + \chi_{\mathbf{77}} \right) q + \mathcal{O}(q^2) \right) \, ,
    \end{aligned}
\end{align}
where $ \chi_{\mathbf{R}} $ is the finite-dimensional $ \mathfrak{g}_2 $ character of the representation $ \mathbf{R} $. The representations $ \mathbf{7} $, $ \mathbf{14} $, $ \mathbf{27} $ and $ \mathbf{77} $ have highest weights $ \omega_1 $, $ \omega_2 $, $ 2\omega_1 $ and $ 3\omega_1 $, respectively. The dual Coxeter labels $ d_j^\vee $ give the levels $ (1,1,2) $ for $ j=0,1,2 $. The exact affine characters are given in Appendix~\ref{app:Lie}. The full partition functions $ \mathcal{W}_j $ in the presence of Wilson surface defects, whose expansion takes the form \eqref{eq:su3-W-expand}, satisfy the blowup equations constructed from the magnetic fluxes \eqref{eq:g2-codim4-flux}. At the $ k $-th order, the blowup equation reads
\begin{align}
    \thtftn{-\nu}{1/2}(3\tau, \lambda+d_j^\vee\epsilon_1) \mathcal{W}_j^{(k)} = \sum_{P_k} (-1)^{n_1 n_2} \thtftn{-\nu}{1/2}(3\tau, \lambda-\mu+d_j^\vee\epsilon_1) \mathcal{A}(n) \mathcal{W}_{j}^{(k_1,N)} Z_{k_2}^{(S)} \, .
\end{align}
The elliptic genera $ \mathcal{W}_j^{(k)} $ can be computed recursively from the bulk elliptic genera and perturbative parts \eqref{eq:g2-affine} by solving the two equations with $ \nu=\pm 1/6 $.

\paragraph{Codimension-2 defect}

Let us consider the codimension-2 defect of the $ \mathfrak{g}_2 $ gauge theory realized by coupling a 4d chiral multiplet in the fundamental representation $ \mathbf{7} $ of the bulk $ \mathfrak{g}_2 $ gauge symmetry, with charge $ +1 $ under the $ \U{1}_x $ flavor symmetry. The defect contribution to the effective prepotential and the perturbative partition function are
\begin{align}
    \mathcal{E}_2 &= \frac{2x}{\epsilon_1} \varphi \, , \quad
    \Psi_{\mathrm{pert}} = \PE\left[ \frac{\sqrt{p_1} \left( e^{2\pi i x} - e^{-2\pi i x} \right)}{(1-p_1)(1-q)} \sum_{j=1}^3 \left( e^{2\pi i a_j} + q e^{-2\pi i a_j} \right) \right] .
\end{align}
Here, we take $ \mathcal{S} $ in \eqref{eq:psi-pert2} to be the set of weights $ \rho $ satisfying $ \langle \rho, \phi \rangle = a_j $. Since $ \sum a_j=0 $, the 1-loop contribution \eqref{eq:codim2-pre-loop} to the prepotential vanishes. The full codimension-2 defect partition function takes the form \eqref{eq:su3-Psi}, and satisfies the blowup equation constructed from the magnetic fluxes \eqref{eq:g2-codim4-flux}. After summing over $ \ell $, the blowup equations at the $ k $-th order read
\begin{align}
    \thtftn{-\nu}{1/2}(3\tau, \lambda+2x) \Psi_k = \sum_{P_k} (-1)^{n_1n_2} \thtftn{-\nu}{1/2}(3\tau, \lambda-\mu+2x) \mathcal{A}(n) \mathcal{D}(n) \Psi_{k_1}^{(N)} Z_{k_2}^{(S)} \, ,
\end{align}
where the defect 1-loop factor is \eqref{eq:so-D}. The two equations with $ \nu=\pm 1/6 $ determine the elliptic genera $ \Psi_k $ in the presence of the defect recursively.

\paragraph{Seiberg--Witten curve}

We construct a quantum difference equation from the defect partition functions. As in section~\ref{sec:so}, the case of $ \so N $ gauge theories, we define the wavefunction $ \psi(x) $ as the NS-limit of the normalized codimension-2 defect partition function in \eqref{eq:so-psi}, including the shift $ \mathcal{E}_2 \to \mathcal{E}_2 - \tau x/(2\epsilon_1) $. We also define $ \chi(x) $ and $ U_j $ by the normalized NS-limits in \eqref{eq:so-codim4-ns}. Similarly to $ \so N $ theories, it is convenient to introduce $ \mathcal{P}_I $, whose string expansions are given by
\begin{align}\label{eq:g2-PI}
    \mathcal{P}_I = Q^{-1} \prod_{j=1}^3 \frac{\theta_I(\tau, a_j)}{\eta(\tau)} + \mathcal{O}(Q^2) \, .
\end{align}
The instanton string corrections are determined from the linear combination of affine $ \mathfrak{g}_2 $ characters:
\begin{align}
    \begin{aligned}\label{eq:g2-P}
        \mathcal{P}_2 &= q^{11/30} H(2\tau) U_0 + q^{-1/30} G(2\tau) U_1 \, , \\
        \mathcal{P}_4 &= q^{-1/120} G(\tau/2) U_0 - q^{11/120} H(\tau/2) U_1 \, ,
    \end{aligned}
\end{align}
where $ G(\tau) $ and $ H(\tau) $ are the Rogers--Ramanujan functions defined in \eqref{eq:ramanujan}. With $ Y f(x) = f(x+\epsilon) $, we propose the quantum SW curve
\begin{align}\label{eq:G2-curve}
    \Big[ Y - (A_2(x) \mathcal{P}_2 - A_4(x) \mathcal{P}_4 ) Y^{-1} - B(x) Y^{-2} \Big] \psi(x) = \chi(x) \psi(x) \, .
\end{align}
The coefficient functions are
\begin{align}
    \begin{aligned} \label{eq:G2-curveAB}
        A_I(x) &= \frac{\theta_1(2x-\epsilon)^2 \theta_1(x-\frac{\epsilon}{2}\pm m) \theta_J(\epsilon) \theta_J(x) \theta_J(x-\epsilon)}{\eta(\tau)^6\theta_3(\tau,0)} \, , \\
        B(x) &= \frac{\theta_1(2x) \theta_1(2x-\epsilon) \theta_1(2x-3\epsilon) \theta_1(2x-4\epsilon) \theta_1(x-\frac{\epsilon}{2}\pm m) \theta_1(x-\frac{3\epsilon}{2}\pm m)}{\eta(\tau)^6 \theta_1(x) \theta_1(x-2\epsilon)} \, ,
    \end{aligned}
\end{align}
which are independent of the dynamical parameters of the theory. In $ A_I(x) $, we take $ J=4 $ if $ I=2 $ and $ J=2 $ if $ I=4 $, and we use the shorthand notation $ \theta_I(u\pm v) = \theta_I(u+v) \theta_I(u-v) $. We have checked this difference equation up to 3-string order. We remark that, as in the $ \so N $ theories, the NS-limit of the codimension-4 defect partition function $ \chi(x) $ can be expressed in terms of the fundamental Wilson surface VEVs $ U_j $ as a linear combination of quadratic products $ U_i U_j $ of the level-one VEVs $ U_0 $ and $ U_1 $, and a linear term in the level-two VEV $ U_2 $, with the coefficients depending only on $ \tau $ and $ x $.

\subsection{Higgsing from \texorpdfstring{$ \so N $}{so(N)} theories to \texorpdfstring{$ \su 3 $}{su(3)} theory}

The theories considered above are connected by Higgs branch flows. These flows provide additional consistency checks of the proposed SW curves. We first consider Higgsing the $ \so N $ gauge theory to the $ \so{N-1} $ theory, by giving a VEV to a scalar field in a hypermultiplet in the vector representation $ \mathbf{N} $. At the level of the partition function, we tune
\begin{align}
    \phi_r \to \phi_{r-1} \, , \quad
    m_f \to -\epsilon_+ \, ,
\end{align}
for even $ N=2r $, where $ m_f $ is the mass parameter of the hypermultiplet in the vector representation that acquires the VEV. For odd $ N=2r+1 $, the Higgsing amounts to tuning
\begin{align}\label{eq:so_odd_higgsing}
    \phi_{r-1} \to \phi_{r-1} + \phi_r \, , \quad
    m_f \to -\epsilon_+ \, .
\end{align}
Under these identifications, the effective prepotential $ \mathcal{E}_0 $ and the perturbative partition function $ Z_{\mathrm{pert}} $ of the $ \so N $ theory reduce to those of the $ \so{N-1} $ theory, up to contributions independent of the dynamical parameters $ \varphi $ and $ \phi_i $ of the theory.\footnote{The $ \so 9 \to \so 8 $ Higgsing reaches only the locus where the masses of the hypermultiplets in the $ \mathbf{8}_s $ and $ \mathbf{8}_c $ representations are equal.\label{footnote:so8}} The blowup equations \eqref{eq:so-blowup} imply that the elliptic genera reduce to those of the $ \so{N-1} $ theory under these parameter identifications. We have also checked these $ \so N \to \so{N-1} $ reductions by comparing the explicitly computed elliptic genera of the parent and daughter theories.

The defect partition functions of the $ \so{N} $ theory also reduce to those of the $ \so{N-1} $ theory under Higgsing. In particular, the perturbative part \eqref{eq:so-X0} of the codimension-4 defect partition function of the $ \so{N} $ theory reduces to its $ \so{N-1} $ counterpart, up to a factor independent of the dynamical parameters. The blowup equations then imply that the full partition function $ \mathcal{X} $ reduces accordingly. The codimension-2 defect partition function $ \Psi $ behaves similarly, since both $ \mathcal{E}_2 $ and the perturbative partition function reduce in the same sense to their counterparts in the daughter theory. For the level-one Wilson surface VEVs, it is convenient to consider the combinations $ \mathcal{P}_I $, whose perturbative parts $ P_I $ defined in \eqref{eq:so-P}, reduce to those of the daughter theory under Higgsing. The linear blowup equations then imply the same reduction for these combinations of level-one Wilson surface VEVs.

We now consider the Higgsing limits of the quantum SW curves \eqref{eq:so-curve}. The coefficient functions $ A_I(x) $ and $ B(x) $ of the $ \so 9 $ curve reduce to those of the $ \so 8 $ curve under the mapping \eqref{eq:so_odd_higgsing}, with the mass identification described in footnote~\ref{footnote:so8}. For $ \so{2r} \to \so{2r-1} $, the coefficients $ A_I(x) $ and $ B(x) $ reduce as follows:
\begin{align}
    \begin{aligned}
        A_I^{\so{2r}} &\to A_I^{\so{2r-1}} \cdot \left( \frac{\theta_1(\tau,x) \theta_1(\tau,x-\epsilon)}{\eta(\tau)^2} \right)^2 \cdot \frac{\eta(\tau)}{\theta_I(\tau, 0)} \, , \\
        B^{\so{2r}} &\to B^{\so{2r-1}} \cdot \frac{\theta_1(\tau, x)^2 \theta_1(\tau, x-\epsilon)^2 \theta_1(\tau,x-2\epsilon)^2}{\eta(\tau)^6} \, .
    \end{aligned}
\end{align}
The extra factors in $ A_I $ cancel when the singlet contributions to $ \chi(x) $, $ \psi(x) $ and $ \mathcal{P}_I $ are included. These singlet contributions arise from the decomposition $ \mathbf{2r} \to \mathbf{2r-1} \oplus \mathbf{1} $ of the $ \so{2r} $ representation into $ \so{2r-1} $ representations, which was omitted in section~\ref{sec:so}.

We next consider the Higgsing from the $ \so 7 $ gauge theory to the $ \mathfrak{g}_2 $ gauge theory, which is realized by giving a VEV to a scalar in one of the spinor hypermultiplets in the $ \so 7 $ theory. At the level of the partition function, this Higgsing amounts to tuning
\begin{align}\label{eq:g2-higgsing}
    \phi_3 \to \phi_1 \, , \quad
    m_2 \to -\epsilon_+ \, .
\end{align}
The remaining mass is identified with the $ \mathfrak{g}_2 $ hypermultiplet mass parameter. The bulk and codimension-4 defect partition functions $ Z $ and $ \mathcal{X} $ reduce to their $ \mathfrak{g}_2 $ counterparts as in the $ \so N $ Higgsings above. For the codimension-2 defect partition function, the differences appear only in the effective prepotential $ \mathcal{E}_2 $ and the perturbative partition function $ \Psi_{\mathrm{pert}} $, reflecting the different choices of the weight set $ \mathcal{S} $ in \eqref{eq:psi-pert2}.

The quantum SW curves are also related by this Higgsing. Applying \eqref{eq:g2-higgsing} in the NS limit to $ B(x) $ of $ \so 7 $ theory gives the $ \mathfrak{g}_2 $ coefficient in \eqref{eq:G2-curveAB}. To match the coefficient of $ Y^{-1} $, we first use the relation
\begin{align}\label{eq:so7-id}
    \theta_2(\tau,0) \mathcal{P}_2^{\so 7}(\tau,\phi) - \theta_3(\tau,0) \mathcal{P}_3^{\so 7}(\tau,\phi) + \theta_4(\tau,0) \mathcal{P}_4^{\so 7}(\tau,\phi) \to 0 \, ,
\end{align}
under the parameter identification \eqref{eq:g2-higgsing}, where $ \mathcal{P}_I^{\so 7} $ is defined from the level-one Wilson surfaces in \eqref{eq:B-P}. At perturbative order, \eqref{eq:so7-id} reduces to a theta function identity \cite{Kharchev:2015}. Assuming the blowup equations, this relation extends to all string orders. The same recursion and matching perturbative seeds give, under \eqref{eq:g2-higgsing},
\begin{align}
    \mathcal{P}_I^{\so 7} \to \mathcal{P}_I^{\mathfrak{g}_2} \, , \quad (I=2,4) \, .
\end{align}
Moreover, the coefficient functions satisfy
\begin{align}
    A_2^{\so 7} - \frac{\theta_2(\tau,0)}{\theta_3(\tau,0)} A_3^{\so 7} \to A_2^{\mathfrak{g}_2} \, , \quad
    A_4^{\so 7} - \frac{\theta_4(\tau,0)}{\theta_3(\tau,0)} A_3^{\so 7} \to -A_4^{\mathfrak{g}_2} \, ,
\end{align}
which follow from the identities of theta functions \cite{Kharchev:2015}. Consequently, the coefficient of $ Y^{-1} $ becomes
\begin{align}
    \sum_{I=2}^4 (-1)^I A_I^{\so 7}(x) \mathcal{P}_I^{\so 7} \to A_2^{\mathfrak{g}_2}(x) \mathcal{P}_2^{\mathfrak{g}_2} - A_4^{\mathfrak{g}_2}(x) \mathcal{P}_4^{\mathfrak{g}_2} \, .
\end{align}
Thus, the quantum SW curve of the $ \so 7 $ theory reduces to that of the $ \mathfrak{g}_2 $ theory under the Higgsing.

Finally, the $ \mathfrak{g}_2 $ theory Higgses to the pure $ \su 3 $ theory by giving a VEV to a scalar in the fundamental hypermultiplet. The fundamental representation $ \mathbf{7} $ of $ \mathfrak{g}_2 $ decomposes as $ \mathbf{3}\oplus \bar{\mathbf{3}} \oplus \mathbf{1} $ under the unbroken $ \su 3 $ gauge symmetry. At the level of the partition function, this Higgsing corresponds to
\begin{align}\label{eq:su3-higgsing}
    \phi_2 \to \phi_1+\phi_2 \, , \quad
    m \to -\epsilon_+ \, .
\end{align}
The bulk prepotential and perturbative partition function reduce to those of the $ \su 3 $ theory, up to contributions independent of the dynamical parameters. The blowup equations imply the corresponding reduction of the elliptic genera. We have also checked this reduction by comparing the explicitly computed elliptic genera of the two theories. The effective prepotential and perturbative partition function of the codimension-2 defect also reduce to those of the $ \mathbf{3}\oplus \bar{\mathbf{3}} $ defect in $ \su 3 $ theory, up to the zero-weight factor. The blowup equations therefore imply the same reduction for the full defect partition function. The codimension-4 defect $ \mathcal{X} $ reduces to 2d fermions in $ \mathbf{3}\oplus \bar{\mathbf{3}} $ representation of $ \su 3 $. Its normalized VEV factorizes in the NS limit, so that
\begin{align}
    \psi^{\mathfrak{g}_2}(x) \to \psi^{\su 3}(x) \, , \quad
    \chi^{\mathfrak{g}_2}(x) \to \chi_+(x) \chi_-(x) \, .
\end{align}
The direct substitution of \eqref{eq:su3-higgsing} into $ B(x) $ of the $ \mathfrak{g}_2 $ theory gives the $ \su 3 $ coefficient in~\eqref{eq:su3-curveAB}. The coefficient of the $ Y^{-1} $ satisfies
\begin{align}
    A_2^{\mathfrak{g}_2}(x) \mathcal{P}_2^{\mathfrak{g}_2} - A_4^{\mathfrak{g}_2}(x) \mathcal{P}_4^{\mathfrak{g}_2} \to A_+(x) \chi_+(x) + A_-(x) \chi_-(x-\epsilon) \, ,
\end{align}
under the Higgsing. At the perturbative level, this identity follows from the theta function identities; the common linear blowup equation extends it to higher string orders. Thus, the Higgsing \eqref{eq:su3-higgsing} reproduces the quantum SW curve of the pure $ \su 3 $ theory, providing a further consistency check of the proposed SW geometry.

\subsection{Comments on \texorpdfstring{$ \mathfrak{su}(3) $}{su(3)} gauge theory with a \texorpdfstring{$ \mathbb{Z}_2 $}{Z2} twist} \label{sec:twist}

We finally comment on the pure $ \su 3 $ gauge theory on a $ -3 $ curve compactified on a circle with a $ \mathbb{Z}_2 $ outer-automorphism twist. This theory is dual to the 5d $ \mathrm{SU}(3) $ gauge theory with Chern-Simons level 9~\cite{Jefferson:2017ahm}, and it can also be obtained by Higgsing the 6d $ \mathfrak{g}_2 $ gauge theory~\cite{Kim:2019dqn, Kim:2021cua}. The outer automorphism exchanges the two nodes of the Dynkin diagram of $ \su 3 $, induces a $ \mathbb{Z}_4 $ grading to the KK modes, and leaves an invariant $ \su 2 $ subalgebra. The effective prepotential and perturbative partition function are expressed as~\cite{Kim:2020hhh}
\begin{align}
    \begin{aligned}
    \epsilon_1 \epsilon_2 \mathcal{E}_0 &= \frac{3}{2} \tau \varphi^2 + 3\varphi \left( -\frac{\epsilon_1^2+\epsilon_2^2}{12} + \phi^2 + \epsilon_+^2 \right) + \frac{4}{3}\phi^3 - \frac{5}{16} \tau \phi^2 - \frac{\epsilon_1^2+\epsilon_2^2}{12} \phi + \epsilon_+^2 \phi \, , \\
    Z_{\mathrm{pert}} &= \PE\left[ -\frac{1+p_1p_2}{(1-p_1)(1-p_2)(1-q)} \bigl( v^2 + (q^{1/4}+q^{3/4})(v+v^{-1}) + q v^{-2} \bigr) \right] \, ,
    \end{aligned}
\end{align}
where $ \varphi $ is the tensor parameter, $ \phi $ is the gauge holonomy for the invariant $ \su 2 $ subalgebra, and $ v=e^{2\pi i \phi} $. For convenience, we make the shift $ \varphi \to \varphi + \frac{5}{48}\tau $ in the prepotential. The blowup equations can be constructed from the magnetic fluxes
\begin{align}
    \ell \in \mathbb{Z} + \frac{h}{3} \, , \quad
    n \in \mathbb{Z} \, , \quad
    B_\tau = 0 \, ,
\end{align}
where $ h=0,1,2 $, and $ \ell $ and $ n $ are the dynamical magnetic fluxes associated with $ \varphi $ and $ \phi $, respectively. The full partition function has the expansion as \eqref{eq:su3-fullZ}, and the $ k $-string elliptic genera $ Z_k $ are computed recursively from the blowup equations.

The elliptic genera $ Z_k $ can also be reproduced by Higgsing the $ \mathfrak{g}_2 $ gauge theory. In addition to the flow \eqref{eq:su3-higgsing}, there is another Higgs branch flow, given by \cite{Kim:2019dqn, Kim:2021cua}
\begin{align}\label{eq:su3-twist-higgs}
    \phi_1 \to \frac{\tau}{2} \, , \quad
    \phi_2 \to \phi + \frac{3}{4} \tau \, , \quad
    m \to \frac{\tau}{2} - \epsilon_+ \, , \quad
    \varphi \to \varphi - \frac{\epsilon_+}{3} \, .
\end{align}
This Higgsing also gives a codimension-4 defect formed by 2d fermions, whose classical and perturbative data are \cite{Kim:2021gyj}
\begin{align}
    \mathcal{E}_1 = 2\varphi \, , \quad
    \mathcal{X}_0 = \prod_{s_1,s_2=\pm 1} \frac{\theta_1(\tau, x + s_1 \phi + s_2 \tau/4)}{\eta(\tau)} \, ,
\end{align}
where we removed the neutral factor in $ \mathcal{X}_0 $. Similarly, the effective prepotential and perturbative partition function of the codimension-2 defect obtained by Higgsing are
\begin{align}
    \mathcal{E}_2 = \frac{2x \varphi}{\epsilon_1} \, , \quad
    \Psi_{\mathrm{pert}} = \PE\left[ \frac{\sqrt{p_1} (e^{2\pi i x}-e^{-2\pi i x})}{(1-p_1)(1-q)} (e^{2\pi i \phi} + e^{-2\pi i \phi}) (q^{-1/4} + q^{5/4}) \right] \, .
\end{align}
Using instead the prescription \eqref{eq:psi-pert1}, the same defect can be written as
\begin{equation}
    \mathcal{E}_2 = \frac{2x \varphi + \phi^2}{\epsilon_1} , \
    \Psi_{\mathrm{pert}} = \PE\left[ \frac{\sqrt{p_1} (e^{2\pi i x}-q e^{-2\pi i x})}{(1-p_1)(1-q)} (e^{2\pi i \phi} + e^{-2\pi i \phi}) (q^{1/4} + q^{-1/4}) \right] .
\end{equation}
The elliptic genera $ \mathcal{X}_k $ and $ \Psi_k $ in the presence of the defects can be computed from the two blowup equations with magnetic fluxes
\begin{align}
    \ell \in \mathbb{Z} \, , \ \mathbb{Z} + 1/3 \, , \quad
    n \in \mathbb{Z} \, , \quad
    B_\tau = B_x = 0 \, .
\end{align}
The resulting elliptic genera agree with Higgsing the partition functions of $ \mathfrak{g}_2 $ theory, up to removing the neutral factor.

The Wilson surface VEVs can be treated in a similar manner. The automorphism twist replaces the usual affine Lie algebra by the twisted algebra $ A_2^{(2)} $. Its two fundamental weights $ \hat{\omega}_0 $ and $ \hat{\omega}_1 $ have levels one and two, respectively. The characters of the corresponding integrable highest weight representations define the perturbative Wilson surface VEVs:
\begin{align}
    \begin{aligned}\label{eq:twist-W}
        W_0^{(0)} &= q^{-1/48} \left( 1 + (v+v^{-1}) q^{1/4} + \mathcal{O}(q^{1/2}) \right) \, , \\
        W_1^{(0)} &= q^{7/60} \left( (v+v^{-1}) + q^{1/4} + \mathcal{O}(q^{1/2}) \right) \, .
    \end{aligned}
\end{align}
The exact forms can be found in Appendix~\ref{app:Lie}. The higher string corrections are again computed from the blowup equations.

Finally, we define the wavefunction $ \psi(x) $ by \eqref{eq:so-psi}, NS-limits of the normalized codimension-4 defect partition function $ \chi(x) $ and Wilson surface VEVs $ U_j $ by \eqref{eq:so-codim4-ns}. The wavefunction then satisfies
\begin{align}\label{eq:su3-twist-sw}
    \Bigl[ Y - A(x) U_0 Y^{-1} - B(x) Y^{-2} \Bigr] \psi(x) = \chi(x) \psi(x) \, ,
\end{align}
where
\begin{align}
    \begin{aligned}
        A(x) &= \frac{q^{1/16} \theta_1(\tau,2x-\epsilon)^2 \theta_4(\tau,\epsilon)}{\eta(\tau)^3} \, , \\
        B(x) &= \frac{q^{1/4} \theta_2(\tau,x) \theta_3(\tau,x) \theta_2(\tau,x-2\epsilon) \theta_3(\tau,x-2\epsilon) \theta_1(\tau,2x-\epsilon) \theta_1(\tau,2x-3\epsilon) }{\eta(\tau)^6} \, .
    \end{aligned}
\end{align}
This curve can also be derived by Higgsing the quantum curve of the $ \mathfrak{g}_2 $ theory and factoring out the neutral factor in the defect contribution arising from the identification \eqref{eq:su3-twist-higgs}. We have checked this difference equation up to 3-string order, supporting its identification as the quantum SW curve of the twisted $ \su 3 $ theory.

\acknowledgments
We would like to thank Hee-Cheol Kim for useful discussions. The research of MK is supported by a KIAS Individual Grant QP097502. SK is supported by National Natural Science Foundation of China grant No. 12250610188 and also by the Brain Pool Program funded by the Ministry of Science and ICT through the National Research Foundation of Korea (RS-2025-25457100). XW is supported by the Fundamental Research Funds for the Central Universities (Grants No. WK2030250140) and the National Natural Science Foundation of China (Grant No.12247103).

\appendix

\section{Special functions} \label{app:special}

We summarize the definitions and properties of the special functions used in this paper. Throughout the paper, we use $ q=e^{2\pi i \tau} $, $ z=e^{2\pi i x} $ with $ \mathrm{Im}\,\tau > 0 $.

\paragraph{Plethystic exponential}
The plethystic exponential is defined as
\begin{align}
    \PE[f(x)] = \exp\left(\sum_{n=1}^\infty \frac{1}{n}f(x^n) \right) \, ,
\end{align}
where $ x $ collectively denotes all multiplicative fugacities.

\paragraph{q-Pochhammer symbol}
The q-Pochhammer symbol is defined as
\begin{align}
    (z;q)_\infty = \prod_{k=0}^\infty (1-zq^k) = \PE\left[ \frac{-z}{1-q} \right] \, , \quad
    (z;q)_n = \frac{(z;q)_\infty}{(z q^n; q)_\infty} \, ,
\end{align}
where $ n \in \mathbb{Z} $.

\paragraph{Dedekind eta function}
The Dedekind eta function is defined by
\begin{align}
    \eta(\tau) = q^{1/24} \prod_{n=1}^\infty (1-q^n) \, .
\end{align}

\paragraph{Theta functions}
The theta function with characteristics $ \alpha, \beta \in \mathbb{R} $ is defined by
\begin{align}
    \theta\genfrac[]{0pt}{1}{\alpha}{\beta} (\tau, x) = \sum_{n \in \mathbb{Z}} q^{\frac{1}{2}(n + \alpha)^2} z^{n+\alpha} e^{2\pi i \beta (n+\alpha)} \, .
\end{align}
The Jacobi theta functions are defined as
\begin{alignat}{2}
    &\theta_1(\tau, x) = -\theta \genfrac[]{0pt}{1}{1/2}{1/2} (\tau, x) \, , \qquad
    &&\theta_2(\tau, x) = \theta \genfrac[]{0pt}{1}{1/2}{0} (\tau, x) \nonumber \\
    &\theta_3(\tau, x) = \theta \genfrac[]{0pt}{1}{0}{0} (\tau, x) \, , \qquad
    &&\theta_4(\tau, x) = \theta \genfrac[]{0pt}{1}{0}{1/2} (\tau, x) \, .
\end{alignat}
For $ m,n \in \mathbb{Z} $, the theta function with characteristics satisfies
\begin{align}
    \thtftn{\alpha}{\beta}(\tau, x+m+n\tau) = e^{2\pi i \alpha m-2\pi i n \beta} q^{-n^2/2} z^{-n} \thtftn{\alpha}{\beta}(\tau,x) \, .
\end{align}
In particular,
\begin{align}\label{eq:tht1-periodic}
    \theta_1(\tau,x+1) = -\theta_1(\tau,x) \, , \quad
    \theta_1(\tau,x+\tau) = -e^{-\pi i \tau - 2\pi i x} \theta_1(\tau,x) \, .
\end{align}
We also use the identities involving theta functions:
\begin{align}
    2\eta(\tau)^3 = \theta_2(\tau,0) \theta_3(\tau,0) \theta_4(\tau,0) \, , \quad
    \prod_{I=1}^4 \theta_I(\tau,x) = \eta(\tau)^3 \theta_1(\tau,2x) \, .
\end{align}
More identities can be found in \cite{Kharchev:2015}.

\paragraph{$ q $-theta function}
We define the $ q $-theta function by
\begin{align}
    \vartheta(\tau, x) = \vartheta(z; q) = (z;q)_\infty (q/z; q)_\infty
    = \frac{z^{1/2} \theta_1(\tau,x)}{iq^{1/12} \eta(\tau)} \, .
\end{align}
It has the following properties:
\begin{align}
    \vartheta(qz; q) = \vartheta(z^{-1}; q) = -z^{-1} \vartheta(z;q) \, .
\end{align}

\paragraph{Elliptic gamma function}
The elliptic gamma function is defined by
\begin{align}
    \Gamma_e(z; p,q)
    := \prod_{m=0}^\infty \prod_{n=0}^\infty \frac{1-p^{m+1}q^{n+1} z^{-1}}{1-p^m q^n z}
    = \PE\left[ \frac{z - pqz^{-1}}{(1-p)(1-q)} \right] \, .
\end{align}
It has the following properties:
\begin{align}
    \Gamma_e(z; p,q) = \frac{1}{\Gamma_e(pq/z; p,q)} \, , \quad
    \Gamma_e(pz; p,q) = \vartheta(z;q) \Gamma_e(z; p,q) \, .
\end{align}

\paragraph{Rogers--Ramanujan functions}
The Rogers--Ramanujan functions are defined by
\begin{align}
    \begin{aligned}\label{eq:ramanujan}
        G(\tau) &= \sum_{n\geq 0} \frac{q^{n^2}}{(q;q)_n} = \prod_{n=0}^\infty \frac{1}{(1-q^{5n+1})(1-q^{5n+4})} \, , \\
        H(\tau) &= \sum_{n\geq 0} \frac{q^{n^2+n}}{(q;q)_n} = \prod_{n=0}^\infty \frac{1}{(1-q^{5n+2})(1-q^{5n+3})} \, .
    \end{aligned}
\end{align}

\section{Lie algebra conventions} \label{app:Lie}

In this appendix, we collect the conventions for the Lie algebras used in this paper. For a simple Lie algebra $ \mathfrak{g} $ of rank $ r $, we denote the simple roots and coroots by $ \alpha_i $ and $ \alpha_i^\vee $, respectively, with $ i=1, \cdots, r $. We normalize the invariant form $ \langle \cdot, \cdot \rangle $ so that long roots have length squared two. The coroots, Cartan matrix $ C_{ij} $ and fundamental weights $ \omega_i $ are defined by
\begin{align}
    \alpha_i^\vee = \frac{2\alpha_i}{\langle \alpha_i, \alpha_i \rangle} \, , \quad
    C_{ij} = \langle  \alpha_i, \alpha_j^\vee \rangle \, , \quad
    \langle \alpha_i^\vee, \omega_j \rangle = \delta_{ij} \, .
\end{align}
We expand Cartan elements $ \phi $ and $ \mu $ in the basis of simple coroots, and a weight $ \lambda $ in the basis of fundamental weights:
\begin{align}
    \phi = \sum_{i=1}^r \phi_i \alpha_i^\vee \, , \quad
    \mu = \sum_{i=1}^r \mu_i \alpha_i^\vee \, , \quad
    \lambda = \sum_{i=1}^r \lambda_i \omega_i \, .
\end{align}
Their scalar products are then given by
\begin{align}
    \langle \phi, \mu \rangle = \sum_{i,j} A_{ij} \phi_i \mu_j \, , \quad
    \langle \lambda, \phi \rangle = \sum_{i} \lambda_i \phi_i \, , \quad
    A_{ij} = \langle \alpha_i^\vee, \alpha_j^\vee \rangle = \frac{2C_{ij}}{\langle \alpha_i, \alpha_i \rangle} \, .
\end{align}
For $ \su N $ and $ \so{2r} $, all roots have length squared two, so $ A_{ij} = C_{ij} $. For $ \so{2r+1} $, the last simple root is short, and $ A_{ij} $ is obtained by doubling the last row of $ C_{ij} $, i.e., $ A_{r-1,r}=A_{r,r-1}=-2 $ and $ A_{rr}=4 $. 
For $ \mathfrak{g}_2 $, the short simple root is $\alpha_1$ and the long simple root is $\alpha_2$. 
This convention gives
\begin{align}
    C = \begin{pmatrix} 2 & -1 \\ -3 & 2 \end{pmatrix} , \quad
    A = \begin{pmatrix} 6 & -3 \\ -3 & 2 \end{pmatrix} .
\end{align}

We normalize the trace $ \Tr $ and quadratic index $ A_{\mathbf{R}} $ of a representation $ \mathbf{R} $ by
\begin{align}
    \Tr F^2 = \frac{1}{h^\vee} \tr_{\mathbf{adj}}F^2 = 2\langle F, F \rangle \, , \quad
    \tr_{\mathbf{R}} F^2 = A_{\mathbf{R}} \Tr F^2 \, ,
\end{align}
where $ h^\vee $ is the dual Coxeter number. The values used in this paper are
\begin{alignat}{2}
    \begin{aligned}
        \su N \quad &: \quad h^\vee = N \, , \quad && A_{\mathbf{N}} = A_{\bar{\mathbf{N}}} = 1/2 \, , \\
        \so N \quad &: \quad h^\vee = N-2 \, , \quad && A_{\mathbf{N}} = 1 \, , \quad A_{\mathbf{S}} = 2^{\lfloor (N-7)/2 \rfloor} \, , \\
        \mathfrak{g}_2 \quad &: \quad h^\vee = 4 \, , \quad && A_{\mathbf{7}} = 1 \, ,
    \end{aligned}
\end{alignat}
where $ \mathbf{S} $ denotes a spinor representation of $ \so N $.

\begin{figure}
    \centering
    \begin{subfigure}[b]{0.45\textwidth}
		\centering
        \includegraphics{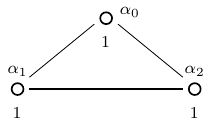}
		\caption{$ \widehat{\mathfrak{su}}(3) $}
		\vspace{3ex}
    \end{subfigure}
    \begin{subfigure}[b]{0.45\textwidth}
		\centering
        \includegraphics{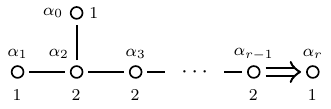}
		\caption{$ \widehat{\mathfrak{so}}(2r+1) $}
		\vspace{3ex}
	\end{subfigure}
    \begin{subfigure}[b]{0.45\textwidth}
		\centering
        \includegraphics{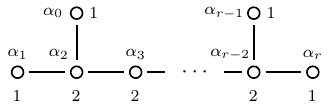}
		\caption{$ \widehat{\mathfrak{so}}(2r) $}
	\end{subfigure}
    \begin{subfigure}[b]{0.45\textwidth}
		\centering
        \includegraphics{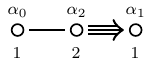}
        \caption{$\widehat{\mathfrak{g}}_2$}
	\end{subfigure}
    \caption{Affine Dynkin diagrams of $ \widehat{\mathfrak{su}}(3) $, $ \widehat{\mathfrak{so}}(N) $ and $\widehat{\mathfrak{g}}_2 $. The numbers next to the nodes are the dual Coxeter labels $ d_i^\vee $.} \label{fig:dynkin}
\end{figure}

For the fundamental Wilson surfaces of untwisted 6d theories, the perturbative VEVs are characters of the untwisted affine algebra $ \hat{\mathfrak{g}} $ associated with the gauge algebra $ \mathfrak{g} $. Figure~\ref{fig:dynkin} fixes the node numbering and dual Coxeter labels $ d_i^\vee $ for the algebras used here. Deleting the affine node $ \alpha_0 $ gives the finite Dynkin diagram of $ \mathfrak{g} $. With this node numbering, we denote the affine fundamental weights by $ \hat{\omega}_i $ for $ i=0,\cdots,r $. An integrable highest weight $ \hat{\lambda} $ and its level $ k $ can be written as
\begin{align}
    \hat{\lambda} = \sum_{i=0}^r n_i \hat{\omega}_i \, , \quad
    k = \sum_{i=0}^r d_i^\vee n_i \, , \quad
    n_i \in \mathbb{N} \, .
\end{align}
The affine character is given by the Weyl-Kac formula \cite{Kac:1990gs}
\begin{align}
    \chi_{\hat{\lambda}}^{\hat{\mathfrak{g}}} &=\frac{\sum_{w\in W} \det(w) \Theta^{\mathfrak{g}}_{w(\lambda+\rho),k+h^\vee}(\tau,\phi)}{\sum_{w\in W} \det(w) \Theta^{\mathfrak{g}}_{w\rho,h^\vee}(\tau,\phi)} \, ,
\end{align}
where $ W $ is the finite Weyl group and $ \rho = \sum_{i=1}^r \omega_i $ is the Weyl vector. For a weight $ \mu $ and a positive integer $ K $, the lattice theta function is
\begin{align}\label{eq:lattice-theta}
    \Theta^{\mathfrak{g}}_{\mu,K}(\tau,\phi) =\sum_{\gamma\in Q^\vee} q^{\frac{\langle K\gamma+\mu,K\gamma+\mu\rangle}{2K}} e^{2\pi i\langle K\gamma+\mu,\phi\rangle} \, ,
\end{align}
with $ Q^\vee = \oplus_i \mathbb{Z} \alpha_i^\vee $ the coroot lattice of $ \mathfrak{g} $. The affine character has the leading expansion
\begin{align}
    \chi_{\hat{\lambda}}^{\hat{\mathfrak{g}}} = q^{h_\lambda - c/24} \left( \chi_{\lambda}^{\mathfrak{g}} + \mathcal{O}(q) \right) \, , \quad
    h_\lambda = \frac{\langle \lambda, \lambda+2\rho \rangle}{2(k+h^\vee)} \, , \quad
    c = \frac{k \dim \mathfrak{g}}{k+h^\vee} \, ,
\end{align}
where $ \chi_{\lambda}^{\mathfrak{g}} $ is the finite-dimensional character with the highest weight $ \lambda $.

For the $ \mathfrak{g}_2 $ theory in section~\ref{sec:g2}, the three characters in \eqref{eq:g2-affine} have highest weights $ \hat{\omega}_0 $, $ \hat{\omega}_1 $, and $ \hat{\omega}_2 $, with levels 1, 1, and 2, respectively. For integers $ u $, $ v $, and a positive integer $ K $, we introduce $ \mu = u \alpha_1 + v \alpha_2 $ and define
\begin{equation}
    \mathcal{T}_{u,v}^{(K)}(\tau,\phi)
    = \Theta^{\mathfrak{g}_2}_{\mu,K}(\tau, \phi) + \Theta^{\mathfrak{g}_2}_{-\mu,K}(\tau,\phi) \, , \quad
    \mathcal{D} = \frac{1}{\eta(\tau)^4} \prod_{j=1}^3 \theta_1(\tau,a_j) \prod_{i<j} \theta_1(\tau,a_i-a_j) \, ,
\end{equation}
where $ (a_1,a_2,a_3) = (\phi_1,\phi_1-\phi_2,-2\phi_1+\phi_2) $ for $ \phi = \phi_1 \alpha_1^\vee + \phi_2 \alpha_2^\vee $, and the $ \mathfrak{g}_2 $ theta function uses $ Q^\vee = 3\alpha_1 \mathbb{Z} + \alpha_2 \mathbb{Z} $. The Weyl-Kac formula then gives
\begin{align}
    \begin{aligned}
        W_0^{(0)} &= \chi_{\hat\omega_0}^{\widehat{\mathfrak{g}}_2}
        =\frac{\mathcal{T}_{5,3}^{(5)}+\mathcal{T}_{4,1}^{(5)}+\mathcal{T}_{1,2}^{(5)} -\mathcal{T}_{5,2}^{(5)}-\mathcal{T}_{4,3}^{(5)}-\mathcal{T}_{1,-1}^{(5)}}{\mathcal{D}} \, , \\
        W_1^{(0)} &= \chi_{\hat\omega_1}^{\widehat{\mathfrak{g}}_2}
        = \frac{\mathcal{T}_{7,4}^{(5)}+\mathcal{T}_{5,1}^{(5)}+\mathcal{T}_{2,3}^{(5)} -\mathcal{T}_{7,3}^{(5)}-\mathcal{T}_{5,4}^{(5)}-\mathcal{T}_{2,-1}^{(5)}}{\mathcal{D}} \, , \\
        W_2^{(0)} &=\chi_{\hat\omega_2}^{\widehat{\mathfrak{g}}_2}
        = \frac{\mathcal{T}_{8,5}^{(6)}+\mathcal{T}_{7,2}^{(6)}+\mathcal{T}_{1,3}^{(6)} -\mathcal{T}_{8,3}^{(6)}-\mathcal{T}_{7,5}^{(6)}-\mathcal{T}_{1,-2}^{(6)}}{\mathcal{D}} \, .
    \end{aligned}
\end{align}
These expressions reproduce the expansions in \eqref{eq:g2-affine}.

\begin{figure}
    \centering
    \includegraphics{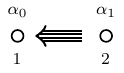}
    \caption{Affine Dynkin diagram and dual Coxeter labels of $ A_2^{(2)} $} \label{fig:twist-dynkin}
\end{figure}

For the $ \su 3 $ theory with an outer-automorphism twist in section~\ref{sec:twist}, the relevant affine algebra is $ A_2^{(2)} $. See \cite{Kim:2020hhh} for its role in twisted compactification and \cite{Kac:1990gs} for the mathematical details. Figure~\ref{fig:twist-dynkin} fixes the node numbering and dual Coxeter labels. The Cartan matrix is given by
\begin{align}
    C=\begin{pmatrix} 2 & -1 \\ -4 & 2 \end{pmatrix} .
\end{align}
It has two fundamental weights $ \hat\omega_0 $ and $ \hat\omega_1 $, which have levels $ d_0^\vee=1 $ and $ d_1^\vee=2 $, respectively. The Weyl-Kac formula gives their characters as
\begin{align}\label{eq:su3-tw-characters}
    \chi_{\hat{\omega}_0} = \frac{\theta_3(\tau/2, \phi)}{\eta(\tau/2)} \, , \qquad
    \chi_{\hat{\omega}_1} = \frac{\Theta_{5,2}(\tau,\phi) - \Theta_{5,-2}(\tau,\phi)}
    {\Theta_{3,1}(\tau,\phi)-\Theta_{3,-1}(\tau,\phi)} \, ,
\end{align}
where
\begin{align}
    \Theta_{\kappa,r}(\tau,\phi) = \sum_{n\in\mathbb Z} q^{(\kappa n+r)^2/(4\kappa)} e^{2\pi i \phi (\kappa n+r)} \, .
\end{align}
The characters reproduce the perturbative Wilson surface VEVs $ W_0^{(0)} $ and $ W_1^{(0)} $ in \eqref{eq:twist-W}.

\bibliographystyle{JHEP}
\bibliography{refs}

\end{document}